\documentclass{JHEP3}

\pdfoutput=1
\usepackage{amsthm,amssymb,amsmath,mathrsfs}
\usepackage{feynmf}
\newcommand{\be}{\begin{equation}}
\newcommand{\ee}{\end{equation}}
\newcommand{\ba}{\begin{eqnarray}}
\newcommand{\ea}{\end{eqnarray}}

\newcommand{\bra}[1]{\left<\,{#1}\,\right|}
\newcommand{\ket}[1]{\left|\,{#1}\,\right>}
\newcommand{\bk}[2]{\left<\,{#1} \mid {#2}\,\right>}
\newcommand{\m}[1]{$\mathop{#1}$}
\newcommand{\scp}[2]{\left\langle\,{#1}\,,\,{#2}\,\right\rangle}
\newcommand{\ru}[2]{\overset{(r)}{#1}{}^{#2}}
\newcommand{\rd}[2]{\overset{(r)}{#1}{}_{#2}}

\newcommand{\nn}{\nonumber\\}
\newcommand{\e}{&=&}

\title{On higher spin interactions with matter}

\author{Xavier Bekaert\\
Laboratoire de Math\'ematiques et Physique Th\'eorique\\
Unit\'e Mixte de Recherche $6083$ du CNRS, F\'ed\'eration Denis Poisson\\
37200 Tours, France\\
\email{bekaert@lmpt.univ-tours.fr}}

\author{Euihun Joung, Jihad Mourad\\
AstroParticule et Cosmologie\\
Unit\'e Mixte de Recherche $7164$ du CNRS\\
Universit\'e Paris VII, B\^atiment Condorcet\\
75205 Paris Cedex 13, France\\
\email{joung@apc.univ-paris7.fr},\email{ mourad@apc.univ-paris7.fr}}

\abstract{Cubic couplings between  a complex scalar field
and a tower of symmetric tensor gauge fields of all ranks are investigated.
A symmetric conserved current, bilinear in the scalar field and
containing $r$ derivatives, is provided for any rank $r\geqslant 1$
and is related to the corresponding rigid symmetry of Klein-Gordon's
Lagrangian.
Following Noether's method, the tensor gauge fields interact with the scalar
field via minimal coupling to the conserved currents.
The corresponding cubic vertex is written in a  compact form
by making use of Weyl's symbols. This enables the explicit computation
of the non-Abelian gauge symmetry group,
the current-current interaction between scalar particles mediated
by any gauge field and the corresponding four-scalar elastic
scattering tree amplitude. The exact summation of these amplitudes
for an infinite tower of gauge fields is possible and several
examples for a definite choice of the coupling constants are provided where the total amplitude exhibits fast
(\textit{e.g.} exponential) fall-off in the high-energy limit.
Nevertheless, the long range interaction potential is dominated by
the exchange of low-spin ($r\leqslant 2$) particles in the low-energy limit.}

\begin{document}

\section{Introduction and summary of results}

The role of higher-spin fields in fundamental interactions is
still unclear. On the one hand, starting from spin two, the
potential  coupling constants have negative mass dimensions
leading to power counting nonrenormalisable theories. On the other
hand, higher spin particles have a crucial role in the softness of
string interactions at high energies; the infinite tower of
massive higher-spin states provides a regularisation in the
ultraviolet. Confronting this fact and the example of Vasiliev's
theory \cite{unfold} (reviewed \textit{e.g.} in
\cite{Vasiliev:2004qz}) with the many no-go theorems
\cite{Weinberg:1964ew,CMHLS,Aragone:1979hx,yesgo} involving a finite
number of massless higher-spin fields suggests that an infinite
collection of higher-spin fields is a necessary ingredient for
building a consistently interacting theory. Furthermore, the algebra of
higher-spin symmetries is expected to have consistent Lorentz covariant
truncations only for gauge fields with spin not greater than two.

Here, we would like to examine the issues of the high energy
behaviour and the gauge symmetries in more details in the
framework of a simple example: the cubic couplings between a
matter scalar field and a collection of higher-spin gauge fields. The
model is consistent from quadratic order in the gauge and matter
fields up to cubic couplings of two scalar and one gauge field.
This model can be used to reliably calculate tree level amplitudes
for the elastic scattering of two scalar particles. It also gives
a hint on the non-Abelian generalization of the gauge algebra, in
our case it is the algebra of unitary operators on
$L^2(\mathbb{R}^n)$ where $n$ is the spacetime dimension.

Let us first describe briefly the model. We start with a free
matter scalar field $\ket\phi$, with the Klein-Gordon action
\be
    S_0[\phi]=-\bra\phi {\hat P}^2+m^2 \ket\phi.
\ee
It gives rise to an
infinite set of conserved Noether currents. Alternatively, the
generating function
\be
\phi(x-q/2)\,\phi^*(x+q/2)=\sum\limits_{r} \frac1{r!}\,\stackrel{(r)}{J}_{\mu_1\ldots\mu_r}\,q^{\mu_1}\ldots\,q^{\mu_r}\,,
\ee
when expanded in the auxiliary variable $u$ gives as
coefficients symmetric conserved currents $J^{(r)}$ which are
improved Noether currents. The higher-spin gauge fields $h^{(r)}$ can also
be grouped in a generating function
$$
h(x,p)\,=\,\sum\limits_{r\geqslant 0}\frac{1}{r!}\,\rd h{\mu_1\ldots\mu_r}(x)\,p^{\mu_1}\ldots p^{\mu_r}
$$
which we interpret as defining
a Hermitian operator $\hat H$ acting on the scalar field.
The currents allow minimal couplings
with the higher-spin fields. The first important remark is that the sum of the
cubic couplings takes the simple form
\be
    S_1[\phi,h]=-\bra\phi \hat H  \ket\phi.
\ee
The precise relationship between the operator $\hat H$
and the generating function $h(x,p)$ for higher-spin field $h^{(r)}$ is that the
latter is the Weyl symbol of the former. Basic facts about Weyl
calculus are recalled in Appendix A.
Denoting $\hat G=\hat P^2+m^2+\hat H$,
 the action $S_0+S_1\,$, which is clearly of the form
$-\bra\phi \hat G \ket\phi\,$, is invariant under the unitary
transformations
\be
    \ket\phi \ \to\ \hat U\ket\phi,
\ee
provided $\hat G$ transforms as
\be
    \hat G\ \to\ \hat U \ \hat G\ {\hat U}^{-1}.
\ee
The second important observation is that this transformation reduces
to lowest order in $\hat H$ to the gauge transformation of
massless higher-spin fields\footnote{ Similar ideas on the link
between Weyl quantisation and Noether couplings between matter and
gauge fields have been pushed forward previously in the context of
conformal higher-spin theory by Segal \cite{Segal:2000ke}. 
Symbol calculus made one of its earliest appearance in the subject of higher spin interactions in the construction of higher-spin (super)algebras \cite{Vasiliev:1986qx}.}. This is shown in Section 3.

We next consider tree level scattering amplitudes which can be
easily calculated in our framework. The model gives rise to cubic
vertices, which together with the propagators of the higher-spin
fields allow the calculation of the tree amplitudes for the scattering of two scalar particles.
The propagators which are suitable for our purposes were found in
\cite{FMS} where no assumption about the vanishing of the double
trace of the fields were made.
 One may ask about the coupling constants of the theory.
In fact, there is an infinite number of them, which are hidden in
the correspondence between $\hat H$ and the higher-spin fields
$h^{(r)}$ or, by a field redefinition, in the kinetic terms of
$h^{(r)}$. We have one coupling constant $\lambda$ with dimension of
length and a collection of dimensionless couplings $a_r$
associated with each spin $r$. In fact all these dimensionless
couplings can be grouped in a generating function $a(z)$
\be
    a(z)=\sum_{r=0}^{\infty}\frac{a_r}{r!}\,z^r.
\ee
We will show that
the tree level amplitude of the two-scalar scattering $\phi\,\phi\,\to\,\phi\,\phi$ and the
non-relativistic potential can both be expressed simply in terms
of this generating function. Its behavior near the origin
determines the static interaction potential and its behavior at
large negative arguments determines the high energy scattering
amplitudes. The explicit expression of the scattering amplitude
turns out to be very simple and is given, in terms of the
Mandelstam variables, by
\be
    A(s,t,u) =
    -\frac{\lambda^{-2}}{t}\ \left[\,
    a\Big(-\frac{\lambda^2}8\left(\sqrt{s}+\sqrt{-u}\,\right)^2\Big)+
    a\Big(-\frac{\lambda^2}8\left(\sqrt{s}-\sqrt{-u}\,\right)^2\Big)-a_0\right]\,.
    \label{sumamplexp1}
\ee
It can be very soft at high energies if the function $a$ is
small for large negative argument. The static potential due to the
exchange of a spin $r$ particle between
two  mass $m$ particles with interdistance $\vec x$ can be deduced and is given by
\be
    \overset{(r)}{V}(\vec x)
    =\frac{a_r}{4\,r!}\,
    \Big(-\frac{(m\lambda)^2}{2}\Big)^{r-1}
    \frac{1}{4\pi\,|\vec x|}\,.
\ee
If $\lambda$ is of the order of the
Planck length and $m$ of the proton mass, then $(m\lambda)\ll 1$ and the potentials for higher
spins are negligible with respect to the Newtonian one provided
the coefficients $a_r$ do not grow fastly with $r$. Unitarity
leads to positive coefficients $a_r$ but otherwise the generating
function is arbitrary within our framework. We expect higher order
consistency to further constrain this function.

\bigskip

The plan of the paper is as follows:
Section \ref{Noethermethod} presents a concise reformulation of the so-called
Noether method for introducing consistent interactions between matter and gauge fields
in terms of various generating functions.
This formalism together with Weyl calculus
is applied in Section \ref{Scalarfield} to the construction of the cubic vertices
that are bilinear in a complex scalar field and linear in a tensor gauge field.
Section \ref{amplitudes} is devoted to the four-scalar elastic
scattering tree amplitude due to the exchange of a single tensor
gauge field. It is  expressed in terms of Chebyshev's or
Gegenbauer's polynomials.
The high-energy behaviour of their sum, corresponding to an infinite tower
of exchanged tensor gauge fields, is discussed in Section \ref{summation}.
The non-relativistic interaction potential is obtained and discussed in Section \ref{applications}.
The paper ends with a short conclusion in Section \ref{conclusion} and several appendices: \footnote{Except in Appendix \ref{Weylquantization}, we set $\hbar=c=1\,$.}
Appendix \ref{Weylquantization} is a brief introduction to the
 formulation of quantum mechanics in terms of Weyl symbols \cite{WW}.
In Appendix \ref{mandelstam}, the Mandelstam variables and various limits of elastic scattering are recalled.
Appendix \ref{Gegenbauer} contains the definitions and some useful formulas for Chebyshev's or Gegenbauer's polynomials.

\section{Generating functions and the Noether method}\label{Noethermethod}

A \emph{symmetric conserved current} of rank $r\geqslant 1$ is a real contravariant
symmetric tensor field $J^{(r)}$ obeying  the conservation law
    \[ \partial_{\mu_1}\ru J{\mu_1\ldots\mu_r}(x)\approx 0\,. \]
where the ``weak equality'' symbol $\approx\,$ stands for ``equal on-mass-shell,''
\emph{i.e.} modulo terms proportional to the Euler-Lagrange equations.
A \emph{generating function of conserved currents} is a real function $J(x,u)$
on phase space which is (i) a formal power series in the ``momenta'' $u_\mu$ and (ii) such that
\begin{equation}
    \left(\frac{\partial}{\partial u_\mu}\,\frac{\partial}{\partial
    x^\mu}\right)\,J(x,u)\approx 0\,. \label{conservationlaw1}
\end{equation}
The terminology follows from the fact that all the coefficients of order $r\geqslant 1$
in the power expansion of the generating function
\begin{equation}
    J(x,u) \,=\, \sum\limits_{r\geqslant 0}\frac{1}{r!}\,\ru J{\mu_1\ldots\mu_r}(x)\,u_{\mu_1}\ldots u_{\mu_r}
\label{j}
\end{equation}
are all symmetric currents which from eq. (\ref{conservationlaw1})
are conserved.

A \emph{symmetric tensor gauge field} of rank $r\geqslant 1$ is a real covariant symmetric tensor field
$h^{(r)}$ whose gauge transformations are \cite{Fronsdal}
\begin{equation}
    \delta_\varepsilon \rd h{\mu_1\ldots\mu_r}(x)\,=
    \,r\,\partial_{(\mu_1} \overset{(r-1)}\varepsilon_{\mu_2\ldots\mu_r)}(x)\,+\,{\cal O}(h)\,,
    \label{gaugetransfo}
\end{equation}
where the gauge parameter $\varepsilon^{(r-1)}$ is a covariant symmetric tensor field
of rank $r-1$ and the round bracket denotes complete symmetrisation with weight one.
For lower ranks $r=1$ or $2\,$, the transformation (\ref{gaugetransfo}) either corresponds
 to the $U(1)$ gauge transformation of the vector ($r=1$) gauge field or to the linearised
 diffeomorphisms of the metric ($r=2$). By comparison with the spin-two case, this formulation
 of higher-spin gauge fields is sometimes called ``metric-like'' (in order to draw the distinction
 with the ``frame-like'' version where the gauge field is not completely symmetric).
A \textit{generating function of gauge fields} is a real function $h(x,v)$ on configuration space
 (i) which is a formal power series in the velocities $v^\mu$ and
 (ii) whose gauge transformations are
\begin{equation}
    \delta_\varepsilon h(x,v)\,=\,\left(v^\mu \frac{\partial}{\partial x^\mu}\right)\,\varepsilon(x,v)\,+\,{\cal O}(h)\,,
    \label{Fronsdalgtransfo}
\end{equation}
where $\varepsilon(x,v)$ is also a formal power series in the velocities.
The nomenclature follows from the fact that all the coefficients of order $r\geqslant 1$ in the power expansion of the generating function
\begin{equation}
	h(x,v)\,=\,\sum\limits_{r\geqslant 0}\frac{1}{r!}\,\rd h{\mu_1\ldots\mu_r}(x)\,v^{\mu_1}\ldots v^{\mu_r}
\label{h}
\end{equation}
are all symmetric tensor gauge fields due to (\ref{Fronsdalgtransfo}) with
$$
\varepsilon(x,v)\,=\,\sum\limits_{r\geqslant 0}\frac{1}{r!}\,\rd\varepsilon{\mu_1\ldots\mu_r}(x)\,v^{\mu_1}\ldots v^{\mu_r}\,.
$$

Of course, in the context of Noether couplings, the ``velocities''
$v^\mu$ and ``momenta'' $u_\nu$ are interpreted as mere auxiliary
variables. Let us introduce a nondegenerate bilinear pairing $\ll \| \gg$
between the generating  functions $J(x,u)$ and $h(x,v)$ on
the configuration and phase spaces respectively,
\begin{equation}
    \ll J\,\|\,h \gg\,\,:=\,\sum\limits_{r\geqslant0}\,
    \int d^nx\ \big<\ \overset{(r)}J(x)\,,\,\overset{(r)}h(x)\ \big>\,,
    \label{pairingsum}
    \end{equation}
where $\scp{J^{(r)}(x)}{h^{(r)}(x)}$ is the contraction between the current and the gauge field:
\be
    \big<\, \overset{(r)}J(x)\,,\,\overset{(r)}h(x)\ \big>=
    \frac{1}{r!}\ \ru J{\mu_1\ldots\mu_r}(x)\,\rd h{\,\mu_1\ldots\mu_r}(x)\,.
    \label{contraction}
\ee
This bilinear form can be written in terms of the generating functions as
\ba
    \ll J\,\|\,h \gg =  \int d^nx\
    \exp\left(\,\frac{{\partial}}{\partial u_{\mu}}\,\frac{{\partial}}{\partial v^{\mu}}\right)
    J(x,u)\,h(x,v)\,\Big|_{u=v=0}\,.
    \label{pairing}
\ea

Let us denote by $\ddagger$ the adjoint operation for the pairing (\ref{pairing}) in the sense that
$$
	\ll J\,\|\,\Hat{\Hat{O}}\,h \gg\,=\,\ll \Hat{\Hat{O}}^\ddagger\, J\,\|\,h \gg\,,
$$
where $\Hat{\Hat{O}}$ is an operator acting on the vector space of functions on configuration space (the double hat stands for ``second quantisation'' in the sense that the operator acts on symbols of ``first quantised'' operators).
Notice that $(v^\mu)^\ddagger={\partial}/{\partial u_\mu}$ and $({\partial}/{\partial x^\mu})^\ddagger=-{\partial}/{\partial x^\mu}$ imply the useful relation
\begin{equation}
	\left(v^\mu \frac{\partial}{\partial x^\mu}\right)^\ddagger\,
	=\,-\,\left(\frac{\partial}{\partial u_\mu}\,\frac{\partial}{\partial x^\mu}\right)\,.
\label{adjoint}
\end{equation}

The \textit{matter action} is a functional $S_0[\,\phi\,]$ of some matter fields collectively denoted by $\phi\,$.
The Euler-Lagrange equations of these matter fields is such that there exists some conserved current $J^{(r)}[\,\phi\,]\,$.
The Noether method for introducing interactions is essentially
the ``minimal'' coupling between a gauge field $h^{(r)}$ and
a conserved current $J^{(r)}[\,\phi\,]$ of the same rank.
Accordingly, the \textit{Noether interaction} between gauge fields
and conserved currents is the functional of both matter and gauge fields
 defined as the pairing between the generating functions
\begin{equation}
    S_1[\,\phi,h\,]\,:=\,\,-\,\ll J\,\|\, h\gg\,.
    \label{Noetherinteraction}
\end{equation}
 Let us assume that there exists a gauge invariant action
$S[\,\phi,h\,]$ whose power expansion in the gauge fields starts as
follows
\begin{equation}
	S\,[\,\phi,h\,]\,=\,S_0[\,\phi\,]\,+\,S_1[\,\phi,h\,]\,+\,S_2[\,\phi,h\,]\,+\,{\cal O}(h^3)\,.
	\label{actionexpansion}
\end{equation}
The  variation of the Noether interaction
(\ref{Noetherinteraction}) under (\ref{Fronsdalgtransfo})
$$
	\delta_\varepsilon S_1[\,\phi,h\,]\,=\,\,-\,\ll J\,\|\, \delta_\varepsilon h\gg\,+\,{\cal O}(h)\,,
$$
is at least of order one in the gauge fields when the equations of motion
for the matter sector are
obeyed,
\begin{equation}
	\delta_\varepsilon S_1[\,\phi,h\,]\,\approx\,{\cal O}(h) \,,
	\label{gtransfoaction}
\end{equation}
because the properties (\ref{conservationlaw1}) and (\ref{adjoint}) imply that
\begin{equation}
	\ll J\,\|\,\Big(v\cdot \frac{\partial}{\partial x}\Big)\,\varepsilon \gg\,
	=\,-\,\ll \Big(\frac{\partial}{\partial u}\cdot \frac{\partial}{\partial x}\Big)\,J\,\|\,\varepsilon \gg\,\,\approx\,0\,.
	\label{gauge/conservation}
\end{equation}
Actually, the crucial property (\ref{gtransfoaction}) works term by term since
$$
	\int d^nx\ \ru J{\mu_1\ldots\mu_r}(x)\ \partial_{\mu_1}\!\overset{(r-1)}\varepsilon_{\mu_2\ldots\mu_r}(x)
	\,=\,-\int d^nx\ \partial_{\mu_1}\ru J{\mu_1\ldots\mu_r}(x)\,\overset{(r-1)}\varepsilon_{\mu_2\ldots\mu_r}(x)\,\approx\,0\,.
$$

The equation (\ref{gtransfoaction}) implies that the action
(\ref{actionexpansion}) is indeed  gauge-invariant at lowest order
in the gauge fields because the terms that are proportional to the
Euler-Lagrange equations $\delta S_0/\delta\phi$ of the matter
sector can be compensated by introducing a gauge transformation
$\delta_\varepsilon\phi$ of the matter fields, independent of the
gauge fields $h$ and linear in the matter fields $\phi\,$, such
that
\begin{equation}
	\delta_\varepsilon \Big(\,S_0[\,\phi\,]+ S_1[\,\phi,h\,]\,\Big)\,=\,{\cal O}(h)\,.
\label{lowestorder}
\end{equation}

\vspace{1mm}

A \textit{Killing tensor field} of rank $r-1\geqslant 0$ on
${\mathbb R}^n$ is a covariant symmetric tensor field
$\overline{\varepsilon}^{(r-1)}$ solution of the generalised
Killing equation
$$
	\partial_{(\mu_1}\!\overset{(r-1)}{\overline{\varepsilon}}_{\mu_2\ldots\mu_r)}(x)=0\,.
$$
A \textit{generating function of Killing fields} is a function $\overline{\varepsilon}(x,v)$
 on configuration space which is (i) a formal power series in the velocities and (ii) such that $\overline{\varepsilon}(x+v\,\tau\,,v)=\overline{\varepsilon}(x,v)$ for any $\tau\,$.
Then the coefficients in the power series
$$
	\overline{\varepsilon}(x,v)\,=\,\sum\limits_{r\geqslant 0}\frac{1}{r!}\,
	\rd{\overline{\varepsilon}}{\mu_1\ldots\mu_r}(x)\,v^{\mu_1}\ldots v^{\mu_r}
$$
are all Killing tensor fields on ${\mathbb R}^n\,$.
The variation (\ref{gaugetransfo}) of the gauge field vanishes if the gauge parameter is a Killing tensor field.
Therefore the corresponding gauge transformation $\delta_{\overline\varepsilon}\phi$ of the matter fields
 is a rigid symmetry of the matter action $S_0[\,\phi\,]\,$:
$$\delta_{\overline\varepsilon}S_0[\,\phi\,]=-\,\delta_{\overline\varepsilon}S_1[\,\phi,h\,]\,\big|_{_{h=0}}=0\,,$$
due to (\ref{lowestorder}) and the fact that
$\delta_\varepsilon\phi$ is independent of the gauge fields. In
turn, this shows that the conserved current $J^{(r)}[\,\phi\,]$
must be equal, on-shell and modulo a trivial conserved current
(sometimes called an ``improvement''), to the Noether current
associated with the latter rigid symmetry of the action
$S_0[\,\phi\,]\,$. A careful look at the one-to-one correspondence
between equivalence classes of rigid symmetries of the action and
conserved currents provided by Noether's theorem (see
\textit{e.g.} the section 2 of \cite{Barnich:2001jy} for a concise
review) allows to prove also the following fact: if the Noether
interaction is translation invariant (\textit{i.e.} the Noether
current does not depend on $x$) then the corresponding rigid
symmetry of the matter field does not depend on $x\,$.

\section{Minimal coupling of a scalar to higher-spin gauge fields}\label{Scalarfield}

\subsection{Conserved current of any rank from scalar action}

Consider a matter sector made of a free complex scalar field $\phi\,$,
of mass square $m^2\geqslant 0\,$,
propagating on Minkowski spacetime with mostly plus metric $\eta_{\,\mu\nu}\,$.
The matter action is the quadratic functional
\begin{equation}
    S_0[\,\phi\,]\,=\,-\int d^nx \left(\,\eta^{\mu\nu}\,\partial_\mu\phi^*(x)\,\partial_\nu\phi(x)+m^2\,\phi^*(x)\,\phi(x)\,\right),
    \label{quadratic}
\end{equation}
which gives an Euler-Lagrange equation: \m{ (\Box -m^2)\phi(x)\approx 0}.
Now consider the following function with an auxiliary variable $q^{\mu}$
\be
    \breve{\rho}(x,q):=\phi^*(x-q/2)\,\phi(x+q/2)\,.
\ee
It obeys  a conservation law,
\begin{eqnarray}
    \left(\eta^{\mu\nu}\, \frac{\partial\ }{\partial q^\mu}\,
    \frac{\partial\ }{\partial x^\nu}\right) \breve{\rho}(x,q) \e
    \phi^*(x-q/2)\,\partial^2\phi(x+q/2)\nonumber\\
    &&-\,\partial^2\phi^*(x-q/2)\,\phi(x+q/2)\,\approx 0\,,
    \label{conservationlaw}
\end{eqnarray}
and can be considered as a generating function for symmetric
conserved currents.  Notice that  eq.(\ref{conservationlaw}) is
similar to eq.(\ref{conservationlaw1}) except that the metric must be
used. \footnote{The Minkowski metric provides an isomorphism between the
tangent and cotangent spaces via the identification
$u^\mu=\eta^{\mu\nu}u_\nu\,$, which induces an isomorphism between
the spaces of functions on the configuration and phase spaces.}
Therefore one finds that a very simple generating function of
conserved currents is $J(x,u)=\breve{\rho}(x,-i\,u)\,$ where
the factor $i$ has been introduced in such a way that the function
is real.  It can be formally written in terms of the wave function $\phi(x)$ as
\begin{equation}
J(x,u)=\phi^*(x+i\,u/2)\,\phi(x-i\,u/2)= \left|\,\phi(x-i\,u/2)\,\right|^2\,.
\label{rhotildde}
\end{equation}
where  reality is manifest.
The condition (\ref{conservationlaw1}) can again be checked by a direct
computation.

Moreover, the Taylor expansion of $J(x,u)$ in power series of $u^{\mu}$ :
\be
    J(x,u)=\sum_{r=0}^{\infty} \ \frac{1}{r!}\ \rd J{\,\mu_1\dots\mu_r}(x)\,u^{\mu_1}\dots u^{\mu_r},
\ee
leads to the explicit expression of the symmetric conserved currents
\begin{equation}
    \rd J{\mu_1\ldots\mu_r}(x)\,=\,\Big(\frac{i}{2}\Big)^r
    \sum_{s=0}^r\ (-1)^s\,\binom{r}{s}\
    \partial_{(\mu_1}\dots\partial_{\mu_s}\phi(x)
    \ \partial_{\mu_{s+1}}\dots\partial_{\mu_r)}\phi^*(x)\,,
    \label{explicitcurrents}
\end{equation}
where all indices of the currents have been lowered because its explicit expression is in terms of derivatives of the scalar field.
These currents are proportional to the ones already introduced in \cite{Berends:1985xx}. Various explicit sets of conserved currents were also provided in \cite{current}.
The symmetric conserved current (\ref{explicitcurrents}) of rank $r$ is bilinear in the scalar field and contains exactly $r$ derivatives.
The currents of any rank are real thus, if the scalar field is real then the odd rank currents are absent due to the factor in front of (\ref{explicitcurrents}).
Notice that the symmetric conserved current of rank two
\begin{equation}
    \overset{(2)}J{}_{\mu\nu}(x) = -\frac14\Big(\partial_\mu\partial_\nu\phi^*(x)\,\phi(x)+\phi^*(x)\,\partial_\mu\partial_\nu\phi(x)
    -2\,\partial_{(\mu}\phi^*(x)\,\,\partial_{\nu)}\phi(x)\Big),
    \label{J2}
\end{equation}
is distinct from the canonical energy-momentum tensor
$$T_{\mu\nu}(x)=\partial_{(\mu}\phi^*(x)\,\partial_{\nu)}\phi(x)-
\frac12\,\eta_{\mu\nu}\left(\,|\partial\phi(x)|^2+m^2|\,\phi(x)|^2\,\right),$$
though, on-shell they differ only from a trivially conserved current since
\begin{equation}
    \overset{(2)}J{}_{\mu\nu}(x)\,\approx\,T_{\mu\nu}(x)\,+\,
    \frac14\,\big(\,\eta_{\mu\nu}\partial^2-\partial_\mu\partial_\nu\big)|\,\phi(x)|^2\,.
    \label{JT}
\end{equation}

\subsection{Noether interactions}

The conserved currents $J^{(r)}$ of eq.(\ref{explicitcurrents})
allow to define Noether interactions between the scalar $\phi$
and gauge fields $h^{(r)}$ as in eq.(\ref{Noetherinteraction}) by
\begin{eqnarray}
&&  S_1[\,\phi,h\,]=-\sum_{r=0}^{\infty}\
    \frac{1}{r!}\int d^nx\,\,\rd h{\mu_1\ldots\mu_r}(x)\, \ru J{ \mu_1\ldots\mu_r}(x)\nonumber
    \\
&&=-\sum_{r=0}^{\infty}
    \Big(\frac{i}{2}\Big)^r\sum_{s=0}^r\frac{(-1)^s}{s!\,(r-s)!}\int d^nx
    \rd h{\mu_1\ldots\mu_r}(x)\,
    \partial^{\mu_1}\dots\partial^{\mu_s}\phi(x)\,\partial^{\mu_{s+1}}
    \dots\partial^{\mu_r}\phi^*(x).\label{cub}
\end{eqnarray}
Similar Noether interactions with scalar field conserved currents were elaborated
in \cite{Berends:1985xx,Fotopoulos,Calimanesti}.
Actually, the above cubic interaction is precisely
of the form mentioned in \cite{GSW},
as can be seen from eq.(\ref{explicitcurrents}).
The sum of terms in the cubic interaction (\ref{cub}) can be
expressed in a concise way exhibiting in a manifest way its symmetries.
In order to do so, we first
introduce the generating function of gauge fields:
\footnote{From (\ref{quadratic}) and (\ref{rhotildde}), it is clear that the generating function $J(x,u)$ has mass dimension $n-2\,$. Thus (\ref{pairing}) shows that $h(x,v)$ has mass dimension $2\,$. Since $v^\mu$ has the dimension of a mass,
the tensor gauge field $h^{(r)}$ of rank $r$ has mass dimension $2-r$\,.
}
\be
    h(x,v)=\sum_{r=0}^{\infty}\ \frac{1}{r!}\ \rd h{\mu_1\dots\mu_r}\ v^{\mu_1}\dots v^{\mu_r},
    \label{gen h}
\ee so that the Noether interaction (\ref{cub}) can be expressed
as  $\ll J||h\gg$ or from
 (\ref{pairing})
 as
 \be
 \int d^nx\
    \exp\left(\,\frac{{\partial}}{\partial u_{\mu}}\,\frac{{\partial}}{\partial v^{\mu}}\right)
    \breve\rho(x,-iu)\,h(x,v)\,\Big|_{u=v=0}.\label{cub2}
    \ee
 Next, we notice that  for any function $f(p)$ and $g(p)$\,:
\be
 \exp\!\left(\frac{\partial\ }{\partial u_\mu}\frac{\partial\ }{\partial v^\mu}\right)\,
 \breve{f}(-iu)\,g(v)\,\Big|_{u=v=0}=\int \frac{d^np}{(2\pi)^n}\ f(p)\, g(p)\,,
\ee
where $\breve{f}(q)$ is the Fourier transform of $f(p)$:
$$
\breve{f}(q^\nu)\,:=\int \frac{d^np}{(2\pi)^{n}}\
f(x^\mu,p_\nu)\,\,e^{{i}{}\, u^\mu\,p_\mu}\,.
$$
This allows us to to express (\ref{cub2}) as

\be
    S_{1}[\,\phi,h\,]=-\frac{1}{(2\pi)^n}\int d^nx\,d^np\
    h(x,p)\ \rho(x,p)\,.
    \label{Noether another}
\ee where $\rho(x,p)$ is the inverse  Fourier transform of
$\breve\rho(x,q)$ over the auxiliary variables $q$. The form
(\ref{Noether another}) of the cubic interaction will be an
essentiel ingredient in exhibiting all the symmetries of the cubic
action in the next subsection.

If we rewrite the expression (\ref{Noether another}) in momentum
space, we can get an even more compact form. By noticing that the
Fourier transform of $\rho(x,p)$ over spacetime variables $x$
reads \ba
    \tilde{\rho}(k,p) \e \int d^nx\,d^nq\ \
     e^{-i\,(k\cdot x+p\cdot q)}\ \phi^*(x-q/2)\,\phi(x+q/2)\nn
     \e \tilde \phi^*(p-k/2)\ \tilde\phi(p+k/2)\,,
\ea
the Noether interaction can be written in a very simple form in terms of $\tilde \phi$\,:
\be
    S_1[\,\phi,h\,]=-\int \frac{d^n\ell}{(2\pi)^{n}}\,
    \frac{d^nk}{(2\pi)^{n}} \ \
    \tilde\phi^*(\ell)\ \tilde h\Big(\ell-k,\frac{k+\ell}2\Big)\ \tilde\phi(k)\,.
    \label{Noether mom}
\ee

\subsection{Weyl formulation}

Using the bra-ket notation for the scalar field \m{\phi(x)=\bk x\phi},
the current generating function $\breve{\rho}(x,q)$ can be written as $\bk{x+q/2}{\phi}\!\bk{\phi}{x-q/2}$.
A very important observation is that, as explained in Appendix \ref{Weylquantization},
this is the Fourier transform over momentum space of the  Wigner
function $\rho(x,p)$ associated to the operator $\ket\phi\!\bra\phi$:
\ba
     \rho(x,p) \e \int d^nq\ e^{-i\,p\cdot q}\,\bk{x+q/2}{\phi}\!\bk{\phi}{x-q/2}\,,
\ea
Thus, the expression of the Noether coupling (\ref{cub2}) can now be simplified
using the Weyl correspondence to
\ba
    S_{1}[\phi,h]
    =
    -\bra\phi \hat{H} \ket\phi\,,
    \label{cubic}
\ea where $\hat{H}:=\mathcal{ W}[\,h\,] $ is the image of the
generating function $h(x,p)$ under the Weyl map $\mathcal{W}$
introduced in  (\ref{Weylmap}). Consequently, the Noether
interaction (\ref{pairing}) defined by the generating functions
(\ref{rhotildde}) and (\ref{gen h}) can be written as the ``mean
value'' over the state $\ket\phi$ of the operator $\hat{H}\,$. The
expression (\ref{Noether mom}) could also have been obtained by
inserting the completeness relations  $\int d^nk/(2\pi)^{n}\ \ket
k\!\bra k =\hat{1}$ between each state in (\ref{cubic}) and apply
the identity (\ref{intkernFtr2}). A cubic interaction with scalar
matter was written in this form by Segal in the somewhat different
context of conformal higher-spin gauge theory \cite{Segal:2000ke}.

By making use of the ``anticommutator ordering'' prescription for
the Weyl map, as explained in the Appendix \ref{Weylquantization},
one finds that the operator $\hat{H}$ starts at lower spin as
\begin{eqnarray}
	\hat{H} \e \overset{(0)}h(\hat X)+
	\frac12\,\Big(\,\hat{P}^\mu\,\overset{(1)} h_\mu(\hat X)+ \overset{(1)}h_\mu(\hat X)\,\hat{P}^\mu\,\Big)\nn
	&&+\,\frac18\,\Big(\,\hat{P}^\mu\,\hat{P}^\nu\, \overset{(2)}h_{\mu\nu}(\hat X)
	+2\,\hat{P}^\mu\,\overset{(2)}h_{\mu\nu}(\hat X)\,\hat{P}^\nu+
	\overset{(2)}h_{\mu\nu}(\hat X)\,\hat{P}^\mu\,\hat{P}^\nu\,\Big)+\ldots
\end{eqnarray}
As one can check, the Noether coupling with the vector gauge field
$h^{(1)}_{\ \mu}$ is the usual electromagnetic coupling. The
Noether coupling with the symmetric tensor gauge field $h^{(2)}_{\
\mu\nu}$ corresponds to the ``minimal'' coupling between a
spin-two gauge field and a scalar density $\phi$ of weight
one-half (minimal in the sense that there is no term containing
the trace $\eta^{\mu\nu}\,h^{(2)}_{\ \mu\nu}$ corresponding to the
linearised volume element in the interaction). This means that
$|\phi|^2$ must be a density of weight one. As can be checked
directly from (\ref{JT}), if the action (\ref{actionexpansion})
includes the rank-two conserved current (\ref{J2}) only, then it
reads
\begin{eqnarray}
	S[\phi,h] = -\int d^nx\sqrt{-g}\,\Big[\,g^{\mu\nu}\,\partial_\mu\Phi^*(x)\,\partial_\nu\Phi(x)
	+\Big(m^2-\frac{R}{8}\,\Big)\,|\,\Phi(x)|^2\,\Big]+{\cal O}({h}^2)\,,\quad
\end{eqnarray}
in terms of the scalar
$\Phi:=(-g)^{-\frac14}\phi\,$, the metric $g_{\mu\nu}:=\eta_{\mu\nu}+h^{(2)}_{\ \mu\nu}+{\cal O}(h^2)$
and the scalar curvature $R\,$.

It is worth emphasising that the cubic interaction $S_1[\phi,h]$
contains $r$ derivatives and grows like the power $r-3+n/2$
of the energy scale by naive dimensional analysis,
so if it involves a tensor field of rank $r> 3-n/2$ then it is not (power-counting) renormalisable.
Notice also that for a real scalar field, the interactions occur with tensor gauge fields of even rank only.

\subsection{Weyl algebra as a non-Abelian gauge symmetry}

Using the braket notation of scalar field where $\partial_{\mu}\phi(x)=i\bra x \hat P_\mu \ket \phi$
and the completeness relation $\hat1=\int d^nx \ket x\!\bra x$,
the Klein-Gordon action (\ref{quadratic}) can be rewritten as
\be
    S_{0}[\phi] =- \bra\phi \hat{P}^2+m^2 \ket\phi,
    \label{free action bk}
\ee
which is (minus) the mean value over the state $\ket\phi$ of
the Hamiltonian (constraint) $\hat{P}^2+m^2$.
The quadratic and cubic functionals (\ref{free action bk}) and (\ref{cubic})
are such that the would-be action (\ref{actionexpansion}) at all orders in the gauge fields starts as
\begin{equation}
    S[\phi,h] =-\bra\phi \hat{G}\ket\phi  + {\cal O}(\phi^3,h^2)\,,
    \label{full}
\end{equation}
where the operator
\begin{equation}
    \hat{G}:=\hat{P}^2+m^2+\hat{H}\,,
    \label{Gop}
\end{equation}
should be interpreted in terms of its Weyl symbol $$g(x,p):=p^2+m^2+h(x,p)\,,$$
as the generating function of the various gauge fields around the Minkowski metric as background.

The linearised gauge transformation (\ref{Fronsdalgtransfo}) of the Weyl symbol $h(x,p)$
can be written as the Poisson bracket between
the function $\varepsilon(x,p)$ and the Weyl symbol of $p^2+m^2$ of a free relativistic particle,
\ba
    \left(p^\mu \frac{\partial\ }{\partial x^\mu}\right)\,\varepsilon(x,p)
    =\frac{1}{2}\,\Big\{\,\varepsilon(x,p)\ ,\ p^2+m^2\,\Big\}_{\textbf{P.B.}}
    =-\frac{i}{2}\,\Big[\,\varepsilon(x,p)\ \overset{\star}{,}\ p^2+m^2\,\Big]\,,
    \label{Fronsdalmometum}
\ea
where $\{\cdot ,\cdot \}_{\textbf{P.B.}}$ is the Poisson bracket and
$[\,\cdot\,\overset\star,\,\cdot\,]$ is the commutator with respect to the Moyal product.
The image of the above formula under the Weyl map leads to
\begin{equation}
    \delta_{\hat E}\,\hat{H}
    =-\frac{i}{2}\,\left[\,\hat E\,,\,\hat{P}^2+m^2\,\right]+{\cal O}(\hat{H})\,,
    \label{infinitesimal}
\end{equation}
where $\hat E$ is the image of $\varepsilon(x,p)$ under the Weyl map.
The variation of the scalar field $\phi$ which guarantees the gauge invariance,
 at lowest order in $h\,$, of the action (\ref{full}) is
\begin{equation}
    \delta_{\hat E}\ket\phi = -\frac{i}{2}\,\hat{E}\,\ket\phi\,,
    \label{infinitesimal2}
\end{equation}
as can be checked directly. At lower orders in the derivative, the
explicit form of the operator $\hat E$ in terms of its
Weyl symbol ${\varepsilon}({x},{p})$
\begin{eqnarray}
    \hat{E} \e \overset{(0)}\varepsilon(\hat X) +
    \underbrace{\frac12\,\Big(\hat{P}^\mu\,\overset{(1)}\varepsilon_\mu(\hat X)+
    \overset{(1)}\varepsilon_\mu(\hat X)\,\hat{P}^\mu\Big)}_{=\
    -i\Big(\overset{(1)}\varepsilon_\mu(\hat X)\,\partial^\mu
    \,+\,\frac12 \,\partial^{\mu}\overset{(1)}\varepsilon_\mu(\hat X)\Big)}
    +\ldots
    \label{oper}
\end{eqnarray}
confirms that following  (\ref{infinitesimal2}) the matter
field $\phi$ transforms
as a scalar density of weight one-half under the (linearised) diffeomorphisms.
The set of all such transformations (\ref{infinitesimal2}) closes under the commutator
and is isomorphic to the Lie algebra of Hermitian operators,
\textit{i.e.} the Lie algebra of quantum observables, corresponding to the Lie group of unitary operators.
If one truncates the tower of gauge fields to the lower-spin sector, 
then there are no further terms represented by dots in (\ref{oper}),
and the Lie algebra of symmetries one is left with is the semidirect sum 
of the local $\mathfrak{u}(1)$ algebra and the algebra of vector fields 
on ${\mathbb R}^n\,$, corresponding to the semidirect product of 
the local $U(1)$ group and the group of diffeomeorphisms.
The form of (\ref{full}) suggests the following finite gauge transformation
\begin{equation}
    \ket\phi\ \longrightarrow\ \hat{U}\ket\phi\,,
    \qquad
    \hat{G}\ \longrightarrow\ \hat{U}\ \hat{G}\ \hat{U}^{-1}\,,
    \label{finite}
\end{equation}
with $\hat{U}:=\exp(-i\,\hat{E}/2)$, because, \textit{at lowest
order} in $\hat{H}\,$, it reproduces the infinitesimal
transformations (\ref{infinitesimal})-(\ref{infinitesimal2}) and
leaves invariant the quadratic form $\bra\phi \hat{G} \ket\phi\,$.
The scalar and gauge fields respectively transform in the
fundamental
 and adjoint representation of the group of unitary operators.
Notice that as long as higher-derivative transformations are allowed
 then the infinite tower of higher-spin fields should be included for consistency of
 the gauge transformations (\ref{finite}) beyond the lowest order.
 The infinitesimal version of (\ref{finite}) written in terms of the Weyl symbols leads
  to the following completion of (\ref{Fronsdalgtransfo})
\begin{eqnarray}
    \delta_\varepsilon \, h(x,p) \e -\frac{i}{2}
    \,\Big[\,\varepsilon(x,p)\ \overset{\star}{,}\ p^2+m^2+h(x,p)\,\Big]
    \label{Moy}\\
    \e \left(\eta^{\mu\nu}\,p_\mu\,\frac{\overrightarrow{\partial}\ }{\partial x^\nu}-
    h(x,p)\ \sin\left[\frac12\left(
    \frac{\overleftarrow{\partial}\ }{\partial x^{\mu}}\,\frac{\overrightarrow{\partial}\ }{\partial p_{\mu}}
    -\frac{\overleftarrow{\partial}\ }{\partial p_{\mu}}\,\frac{\overrightarrow{\partial}\ }{\partial x^{\mu}}
    \right)\right]\,\right)\varepsilon(x,p),\nonumber
\end{eqnarray}
where we made use of (\ref{Moyalcommutator}) and
(\ref{Fronsdalmometum}). Such a deformation of the higher-spin
gauge transformations was already advocated in \cite{Segal:2000ke,G,XB}.\footnote{This deformation was already implicit in \cite{V} in the sense that (\ref{Moy}) should arise after the elimination of the auxiliary variables $y\,$.} Notice
that, in general, the Moyal bracket contains a non-vanishing
contribution at $p_{\mu}=0$ which corresponds in (\ref{Moy}) to a
gauge transformation of a tensor field of rank $r=0\,$. Hence, it
might be necessary for the consistency of the non-Abelian gauge
transformations (\ref{Moy}) to include a scalar field $h^{(0)}$ in
the tower of gauge fields \footnote{A scalar field is also
necessary for consistency of Vasilev's unfolded equations. It should be stressed that the transformation (\ref{Moy}) of the gauge scalar field $h^{(0)}$ is distinct from the transformation (\ref{infinitesimal2}) of the matter scalar field $\phi\,$. }.

The Weyl symbol $\overline{\varepsilon}(x,p)$ of an operator
$\hat{\overline{E}}$ commuting with $\hat{P}^2+m^2$ is a generating
function of Killing fields, as can be easily seen  from
(\ref{Fronsdalmometum}). This is in agreement with the facts that
if \m{[\,\,\hat{\overline{E}}\,,\,\hat{P}^2+m^2\,]=0}
then the corresponding transformation (\ref{finite}),
\begin{equation}
    \ket\phi \ \longrightarrow\
    \exp(-i\,\hat{\overline E}\,) \ket\phi,
    \label{rigid}
\end{equation}
is obviously a symmetry of the Klein-Gordon action
(\ref{quadratic}). It is very tempting to conjecture that the full
action (\ref{full}) should be interpreted as arising from the
gauging of the rigid symmetries (\ref{rigid}) of the free scalar
field, which generalise the $U(1)$ and Poincar\'e symmetries, so
the local symmetries (\ref{finite}) generalise the local $U(1)$
and diffeomorphisms. The rigid higher-derivative symmetries which
are generated by a function ${\overline{\varepsilon}}(p)$
independent of the position and which thereby generalise the phase
shifts and translations were introduced in \cite{Berends:1985xx} and further developed in \cite{current}.
The corresponding infinitesimal symmetries are the most general
rigid linear symmetry transformations of a free scalar field which
are independent of the coordinates and compatible with locality
(in the sense that the order of the differential operators is
finite). The group of unitary operators was already advertised in
\cite{Calimanesti} as the symmetry group arising from the gauging
of these rigid higher-derivative symmetries.

Notice that the conserved currents (\ref{explicitcurrents}) are
indeed equivalent to the Noether currents for the latter
symmetries, as follows from Noether's first theorem or as can be
checked by direct computation. This correspondence also implies
that the Noether interaction considered here is the most general
one (up to equivalence) between one gauge field and two free
scalars that induces a gauge transformation of the scalar field
 and is compatible with locality and Poincar\'e symmetry.
In the case of a real scalar field, the Lie algebra and group of
gauge symmetries would have to be replaced by, respectively, the
algebra of symmetric operators and the group of orthogonal
operators. The former construction goes along the same line for a
scalar field taking values in an internal finite-dimensional
space, \textit{i.e.} for a multiplet of scalar fields.

\section{Tree-level higher-spin exchange amplitudes}\label{amplitudes}

\subsection{Feynman rules}

\paragraph{Vertex}

The cubic vertex between two scalar fields $\phi$ and a gauge field $h^{(r)}$
takes a simple form in momentum space in terms of the Fourier transforms of
fields, $\tilde\phi\,$. Indeed, from eq.(\ref{Noether mom}),
the Noether interaction between $\phi$ and $h^{(r)}$ is given by
\ba
    S_{1}[\,\phi,h^{(r)}\,]\e -\int \frac{d^n\ell}{(2\pi)^n}\,\frac{d^nk}{(2\pi)^n}\ \
    \tilde\phi^*(\ell)\ \tilde\phi(k)\ \rd{\tilde h}{\mu_1\dots\mu_r}(\ell-k)\times \nn
    &&\qquad \qquad \times\ \frac{1}{r!}\, \left(\frac{k^{\mu_1}+\ell^{\mu_1}}2\right)
    \dots \left(\frac{k^{\mu_r}+\ell^{\mu_r}}2\right)\,.
    \label{inter mom}
\ea
The corresponding cubic vertices are
\vspace{2mm}
\begin{fmffile}{vertex}
\ba
    \ru{\mathcal{V}}{\,\mu_1\dots\mu_r}(k,\ell)\e
    \parbox{30mm}{
    \begin{fmfgraph*}(30,20)
    \fmfleft{i1,i2}
    \fmfright{o}
    \fmf{fermion,label=$\phi$}{i1,v,i2}
    \fmf{photon,label=$h^{(r)}$}{v,o}
    \fmflabel{$k$}{i1}
    \fmflabel{$\ell$}{i2}
    \fmflabel{$k-\ell$}{o}
    \fmfdot{v}
    \end{fmfgraph*}}
    \nn\
\nn\
\nn
    \e-\,\frac{1}{r!}\,
    \left(\frac{k^{\mu_1}+\ell^{\mu_1}}2\right)
    \dots \left(\frac{k^{\mu_r}+\ell^{\mu_r}}2\right)\,.
    \label{vertex r}
\ea
\end{fmffile}
If the scalar field is real then one can insert the relation $\tilde{\phi}^*(-k)=\phi(k)$ in (\ref{inter mom})
and recover the fact that cubic vertices for odd $r$ are absent in such case.

\paragraph{Propagators}

The propagator with respect to the scalar field $\phi$ is easily determined from the kinetic term in (\ref{free action bk}) and is given by
\begin{fmffile}{propagators}
\ba
    \mathcal{D}(p)\e
     \parbox{30mm}{
    \begin{fmfgraph*}(30,10)
    \fmfleft{i}
    \fmfright{o}
    \fmf{fermion}{i,o}
    \end{fmfgraph*}}
    \nn
    \e \,\frac1{p^2+m^2}\,.
\ea
\end{fmffile}

The current-current interaction which determines the propagator $\mathcal{P}^{(r)}/p^2$
for spin $r$ exchange was determined in \cite{FMS},
and the amplitudes were shown to propagate the correct numbers of on-shell degrees of freedom,
exactly like in Fronsdal's  formulation, even though the currents
involved were not doubly traceless.
The contraction of the propagator residue $\mathcal{P}^{(r)}$ 
with two conserved currents $J_1^{(r)}$ and $J_2^{(r)}$ is given by
\be
    \big<\  \overset{(r)}J_1\,,\,\overset{(r)}{\mathcal{P}}\ \overset{(r)} J_2\ \big>=
    \sum_{m=0}^{\lfloor r/2 \rfloor} \,
    \frac{1}{2^{2m}\,m!\,(3-\frac{n}{2}-r)_m}\
    \big<\ \overset{(r)}J_1{}^{[m]}\,,\, \overset{(r)}J_2{}^{[m]}\ \big>\,,
    \label{fmsfin}
\ee
where $J^{[m]}$ denotes the $m$-th trace of the external current,
 $(a)_m$ denotes the $m$-th Pochhammer symbol of $a$\,: $(a)_{m}=\Gamma(a+m)/\Gamma(a)$\,.

This amplitude corresponds to
a kinetic term for the spin $r$ fields which is canonically
normalised that is of the form
$S_{\text{kin}}[\,h^{(r)}\,]=\frac12< h^{(r)}\, |\,\Box h^{(r)} >+\dots$
However, our gauge fields $h^{(r)}$ are not canonically normalised and
the kinetic terms compatible, to lowest order, with our symmetries have the form
\be
    S_{\text{kin}}[\,h^{(r)}\,]=\frac{\lambda^{6-n-2r}}{2a_r}\
    \big<\overset{(r)}h\, |\,\Box \overset{(r)}h \big>  +\dots
\ee
with $\lambda$ a length parameter and $a_r$
real strictly positive dimensionless
parameters.
Thus, the propagator with respect to our $h^{(r)}$ is given, in a gauge \textit{\`a la} Feynman and de Donder, by
\begin{fmffile}{propagator}
\ba
    \rd{\mathcal{D}}{\mu_1\dots\mu_r\,|\,\nu_1\dots\nu_r}(k)\e \quad\quad\quad\quad
     \parbox{30mm}{
    \begin{fmfgraph*}(30,10)
    \fmfleft{i}
    \fmfright{o}
    \fmf{photon}{i,o}
    \fmflabel{$\mu_1\dots\mu_r$}{i}
    \fmflabel{$\nu_1\dots\nu_r$}{o}
    \end{fmfgraph*}}
    \nn
    \e\frac{a_r\,\lambda^{n-6+2r}}{k^2}\ \rd{\mathcal{P}}{\mu_1\dots\mu_r\,
    |\,\nu_1\dots\nu_r}\,.\label{propa}
\ea
\end{fmffile}

\subsection{Tree-level amplitude}

We consider the following diagram where
two scalar particles of same charge
exchange one gauge particle of rank $r$ in $t$-channel.
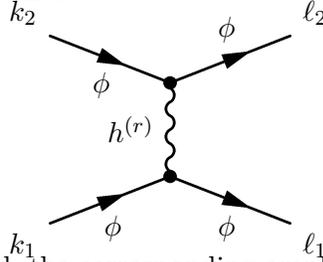
\vspace{2mm}
\begin{center}
 \begin{fmffile}{tree}
  \begin{fmfgraph*}(40,25)
  \fmfleft{i1,i2}
   \fmfright{o1,o2}
   \fmf{fermion,label=$\phi$}{i1,v1,o1}
   \fmf{fermion,label=$\phi$}{i2,v2,o2}
   \fmf{photon,label=$h^{(r)}$}{v1,v2}
   \fmflabel{$k_1$}{i1}
   \fmflabel{$k_2$}{i2}
   \fmflabel{$\ell_1$}{o1}
   \fmflabel{$\ell_2$}{o2}
   \fmfdot{v1,v2}
  \end{fmfgraph*}
 \end{fmffile}
\end{center}
Since vertices $\mathcal{V}^{(r)}$ are conserved,
the corresponding amplitude is given by
\be
    \overset{(r)}{A}\big(\,\phi(k_1)\,\phi(k_2)\,\to\,\phi(\ell_1)\,\phi(\ell_2)\,\big) =
    \Big<\ \overset{(r)}{\mathcal{V}}(k_1,\ell_1)\,,\,
    \overset{(r)}{\mathcal{D}}(k_1-\ell_1)\
    \overset{(r)}{\mathcal{V}}(k_2,\ell_2)\ \Big>\,,
\ee
where the contraction notation (\ref{contraction}) was used and
the momentum conservation $k_1+k_2=\ell_1+\ell_2$ is assumed.
We recall that the propagator $\cal D$
is given in (\ref{propa}) and (\ref{fmsfin}).
Using eq.(\ref{vertex r}), the contraction between two $m$-th traces of vertex is given by
\ba
    &&\Big<\ \overset{(r)}{\mathcal{V}}{}^{[m]}(k_1,\ell_1)\,,\,
    \overset{(r)}{\mathcal{V}}{}^{[m]}(k_2,\ell_2)\ \Big>=\nn
    &&\quad =\,\frac{1}{(r-2m)!}
    \left[\frac{k_1+\ell_1}2 \cdot \frac{k_2+\ell_2}2 \right]^{r-2m}
    \left[\left(\frac{k_1+\ell_1}2\right)^2\, \left(\frac{k_2+\ell_2}2\right)^2\right]^m\,.
\ea
By making use of the above result and the Mandelstam variables $s$, $t$ and $u$
(see Appendix \ref{mandelstam} for more details)\,:
\ba
    &&(k_1+\ell_1)\cdot(k_2+\ell_2)=-(s-u)\,,\nn
    &&(k_1+\ell_1)^2=(k_2+\ell_2)^2=-(s+u)\,,\nn
    &&(k_1-\ell_1)^2=(k_2-\ell_2)^2=-t\,,
\ea
the amplitude can be written as \footnote{
    The case $r=0$ corresponds to the exchange of a scalar ``gauge''
    field $h^{(0)}$ and so is slightly less natural from a physical
    perspective than cases $r\geqslant 1\,$.}
\be
    \overset{(r)}{A}(s,t,u)
    =\, -\frac{\lambda^{n-6}}{t}\ a_r\, \left(-\frac{\lambda^2}{4}\right)^{r}\
    \sum^{\lfloor r/2 \rfloor}_{m= 0}\,
    \frac{(s-u)^{r-2m}\,(s+u)^{2m}}{2^{2m}\,m!\,(r-2m)!\,(3-\frac{n}2-r)_m}\,.
    \label{ramplitude}
\ee

In four-dimensional spacetime ($n=4$) and for $r\geqslant 1\,$,
the sum (\ref{ramplitude}) can be expressed in terms of Chebyshev
polynomials of the first kind (\ref{Tdefinition}) as \be
    \overset{(r)}{A}(s,t,u)
    =-\frac{\lambda^{-2}}{t}\ a_r\,
    \left(-\,\frac{\lambda^2}8\, (s+u)\right)^{r}\,\frac2{r!}\
    T_r\left(\frac{s-u}{s+u}\right)\,.
\ee
In higher dimensions ($n\geqslant 5$),
the sum (\ref{ramplitude}) can be expressed in terms of Gegenbauer
polynomials  (\ref{Gdefinition}) as
 \be
    \overset{(r)}{A}(s,t,u) =
    -\frac{\lambda^{n-6}}{t}\ a_r\,
     \left(-\,\frac{\lambda^2}8\, (s+u)\right)^{r}\,
    \frac1{(\frac{n}2-2)_r}\
    C^{\frac{n}2-2}_r\left(\frac{s-u}{s+u}\right)\,.
    \label{rampl}
\ee
Notice that in $n=5$ dimensions, the Gegenbauer polynomial 
in (\ref{rampl}) essentially becomes a Legendre polynomial.
These amplitudes have a pole when $t$ is equal to the squared mass
of an exchanged particle. Thus for massless mediators $t$ must be
different from zero, \textit{i.e.} the scattering angle
$\theta\neq 0$ modulo $\pi\,$.\footnote{ The scattering angle
$\theta$ in the center-of-mass system is determined by
    \begin{equation}
    \sin^2(\theta/2)= -t/(s-4\,m^2)\,,\qquad
    \cos^2(\theta/2)= -u/(s-4\,m^2)\,.
    \label{thetangle}
    \end{equation}
Since $s\geqslant 4m^2\,$, one should have $t\leqslant 0$ and $u\leqslant 0\,$.
See Appendix \ref{mandelstam} for more details.}

For bosons, the total amplitude for the scattering process 
$\phi(k_1)\,\phi(k_2)\,\to\,\phi(\ell_1)\,\phi(\ell_2)$ contains the sum of the $t$ and $u$ channel amplitude:
\vspace{2mm}
\begin{fmffile}{feyn1}
\ba
    \overset{(r)}A_{\text{total}}(\phi\,\phi\,\to\,\phi\,\phi)\e
    \parbox{30mm}{
    \begin{fmfgraph*}(30,20)
    \fmfleft{i1,i2}
    \fmfright{o1,o2}
    \fmf{fermion}{i1,v1,o1}
    \fmf{fermion}{i2,v2,o2}
    \fmf{photon}{v1,v2}
    \fmflabel{$k_1$}{i1}
    \fmflabel{$k_2$}{i2}
    \fmflabel{$\ell_1$}{o1}
    \fmflabel{$\ell_2$}{o2}
    \end{fmfgraph*}}
    \quad+\quad
    \parbox{30mm}{
    \begin{fmfgraph*}(30,20)
    \fmfleft{i1,i2}
    \fmfright{o1,o2}
    \fmf{fermion}{i1,v1}
    \fmf{fermion}{i2,v2}
    \fmf{photon}{v1,v2}
    \fmf{phantom}{v1,o1}
    \fmf{phantom}{v2,o2}
    \fmf{fermion,tension=0}{v1,o2}
    \fmf{fermion,tension=0}{v2,o1}
    \fmflabel{$k_1$}{i1}
    \fmflabel{$k_2$}{i2}
    \fmflabel{$\ell_1$}{o1}
    \fmflabel{$\ell_2$}{o2}
    \end{fmfgraph*}}
    \nn\
\nn\
\nn
    \e \qquad \overset{(r)}A(s,t,u)\qquad\quad +\quad\qquad \overset{(r)}A(s,u,t)\,.
\ea
\end{fmffile}

The diagrams for the scattering $\phi(k_1)\,\bar\phi(k_2)\,\to\,\phi(\ell_1)\,\bar\phi(\ell_2)$
can be obtained from $A^{(r)}$ by a \textit{crossing} symmetry:
\vspace{2mm}
\begin{fmffile}{feyn2}
\ba
    \parbox{30mm}{
    \begin{fmfgraph*}(30,20)
    \fmfleft{i1,i2}
    \fmfright{o1,o2}
    \fmf{fermion}{i1,v1,o1}
    \fmf{fermion}{o2,v2,i2}
    \fmf{photon}{v1,v2}
    \fmflabel{$k_1$}{i1}
    \fmflabel{$-k_2$}{i2}
    \fmflabel{$\ell_1$}{o1}
    \fmflabel{$-\ell_2$}{o2}
    \end{fmfgraph*}}
    = \overset{(r)}A(u,t,s)\,,
    \quad \quad
    \parbox{30mm}{
    \begin{fmfgraph*}(30,20)
    \fmfleft{i1,i2}
    \fmfright{o1,o2}
    \fmf{fermion}{i1,v1}
    \fmf{fermion}{v1,i2}
    \fmf{photon}{v1,v2}
    \fmf{fermion}{v2,o1}
    \fmf{fermion}{o2,v2}
    \fmflabel{$k_1$}{i1}
    \fmflabel{$-k_2$}{i2}
    \fmflabel{$\ell_1$}{o1}
    \fmflabel{$-\ell_2$}{o2}
    \end{fmfgraph*}}
    =\overset{(r)}A(u,s,t)\,.\ \nn
 \ea
 \end{fmffile}
The parity properties of Gegenbauer and Chebyshev polynomials are such that
\be
    \overset{(r)}{A}(u,t,s)
    =(-1)^r\ \overset{(r)}{A}(s,t,u)\,,
\ee
which is consistent with crossing ``symmetry.''
For instance, if the scalar field is real then the amplitude is
a symmetric function of $s$ and $u\,$.

\bigskip

If $\lambda$ is thought as Planck's length and $m$ as, say,
the proton mass, then $\lambda\,m\approx 10^{-19}\ll 1\,$.
The high-energy regime must now be understood as $s\gg\lambda^{-2}\gg m^2\,$.
In the Regge limit,
the $t$-channel tree-level amplitudes behave as
$$
\overset{(r)}{A}(s,t,u)
\sim -\ \frac{\lambda^{n-6}}{t}\ \frac{a_r}{r!}\ \Big(-\frac{\lambda^2}{2}\,s\Big)^{r}\,,
$$
and for fixed scattering angle $\theta$ in $n=4$ as
$$
\overset{(r)}{A}(s,t,u)\sim
-\ \frac{1}{4}\ \frac{a_r}{r!}\ \Big(-\frac{\lambda^2}{8}\,\sin^2(\theta/2)\,s\Big)^{r-1}\
T_r\!\left(\frac{1+\cos^2(\theta/2)}{\sin^2(\theta/2)}\right)\,.
$$
As one can see, in the latter limit each amplitude grows as the $(r-1)$-th power of the large $s\,$,
so it goes to a constant when $r=1$ and it diverges for spin $r\geqslant 2\,$.
This is another signal of the well-known fact that the corresponding interactions are or not (power-counting) renormalisable.

\section{Summation of tree amplitudes and high-energy behaviour}\label{summation}

In the present section, the main focus is on spacetime dimension
$n=4$ for obvious physical reasons (and because the  case
$n\geqslant 5$ goes exactly along the same lines). For the process
$\phi\, \phi\,\to\,\phi\, \phi$, the sum of the $t$-channel
tree-level amplitudes including all exchanged particles is \be
    A(s,t,u) =\sum_{r\geqslant 0}\ \overset{(r)}{A}(s,t,u)
    =-\frac{\lambda^{-2}}{t}\ \left[a_0+\sum_{r\geqslant 1}\ a_r\,
    \left(-\frac{\lambda^2}8(s+u)\right)^{r}\,\frac{2}{r!}\ T_r\!\left(\frac{s-u}{s+u}\right)\right]\,.
    \label{sumampl}
\ee
Let us denote by $a(z)$ the generating function of the coefficients $a_r$($\geqslant 0$), in the sense that
\be
a(z)=\sum\limits_{r\geqslant 0}\ \frac{a_r}{r!}\ z^r\,.
\label{pow ser}
\ee
Using the identity (\ref{Chebypower}), the sum (\ref{sumampl})
over $r$ can be explicitly performed and gives
\be
    A(s,t,u)=\,-\frac{\lambda^{-2}}{t}\
     \left[\,a\Big(-\frac{\lambda^2}8\left(\sqrt{s}+\sqrt{-u}\,\right)^2\Big)+
     a\Big(-\frac{\lambda^2}8\left(\sqrt{s}-\sqrt{-u}\,\right)^2\Big)-a_0\right]\,.
    \label{sumamplexp}
\ee

In the high-energy regimes $s\gg \lambda^{-2}\gg m^2\,$, the $t$-channel tree amplitude behaves in the Regge limit as
\be
    -\frac{\lambda^{-2}}{t}\,a\Big(-\frac{\lambda^2}{2}\,s\Big)\,,
    \label{ahighenergy}
\ee
and in the fixed scattering angle limit as
\be
    \frac{\lambda^{-2}}{\sin^2(\theta/2)\,s}\,
    \left[\,a\Big(-\frac{\lambda^2}8\,\big[1-\cos(\theta/2)\big]^2\,s\Big)
    +a\Big(-\frac{\lambda^2}8\,\big[1+\cos(\theta/2)\big]^2\,s\Big)-a_0\right],
    \label{ahighentheta}
\ee
which formally reproduces the behaviour (\ref{ahighenergy}) in the limit $\theta\to 0$ with fixed (but large) $s\,$.

\subsection{Simplest examples}

We first consider
the simplest choice of coefficients: $a_r=1$ for all
$r\geqslant0\,$. Hence $a(z)=e^z$ so that the $t$-channel
amplitude is equal to \be A(s,t,u) =
-\frac{\lambda^{-2}}{t}\,\left[\,
2\,\exp\!\Big(-\frac{\lambda^2}8\,(s-u)\Big)\,
\cosh\!\Big(\,\frac{\lambda^2}{4}\,\sqrt{-su}\,\Big)-1\,\right]\,,\label{amp1}
\ee and decreases exponentially in the Regge limit,
$$
A(s,t,u)
\sim -\frac{\lambda^{-2}}{t}\ \exp\!\left(-\frac{\lambda^2}{2}\,s\right)
$$
in agreement with (\ref{ahighenergy}).
Next, in order to cancel the constant contribution in the brackets of
(\ref{amp1}) we consider
another choice of coefficients  $a_0=2$ and $a_r=1$ for all $r\geqslant1\,$.
Hence $a(z)=e^z+1$ and the $t$-channel amplitude is equal to
$$
A(s,t,u) =
-\frac{2\,\lambda^{-2}}{t}\,
\exp\!\Big(-\frac{\lambda^2}8\,(s-u)\Big)\,
\cosh\!\Big(\,\frac{\lambda^2}{4}\,\sqrt{-su}\,\Big)\,,
$$
and falls-off exponentially for large $s$ but fixed scattering angle $\theta\neq 0$
$$
A(s,t,u) \sim \frac{\lambda^{-2}}{\sin^2(\theta/2)\ s}\,\,
\exp\Big(-\frac{\lambda^2}8\,\big[1-\cos(\theta/2) \big]^2\ s\,\Big)
$$
as can be checked directly or from
(\ref{ahighentheta}).
However, the $t$-channel tree-level
scattering amplitude of the process
$\phi\,\bar\phi\,\to\,\phi\,\bar\phi$ grows
exponentially.

\subsection{General discussion}

Let $a(z)$ be the real function defined by the power series (\ref{pow ser})
with non-negative coefficients $a_r\geqslant 0\,$.
Let us assume that the function is holomorphic on the complex plane 
except a set of isolated poles (\textit{i.e.} it is meromorphic) which does not contain the origin.
More concretely, the function $a(z)$ is analytic inside the disk of convergence 
of the power series $\sum_{r\geqslant 0}\ \frac{a_r}{r!}\ z^r$ around 
the origin $z=0$ and it is defined outside the radius of convergence by analytic continuation.

The poles of the corresponding $t$-channel tree-level amplitude
for the exchange of an infinite tower of tensor gauge fields
between two scalar particles might be interpreted, effectively, as
the exchange of some massive particles. This amplitude goes to
zero in the Regge limit if and only if $z=-\infty$ is a zero of
$a(z)$, as can be seen from (\ref{ahighenergy}). Moreover, at any
fixed scattering angle $\theta\neq 0$ (modulo $\pi$), the
high-energy limit of the $t$-channel tree-level amplitude goes to
zero if $a(z)$ goes to a constant at $z=-\infty\,$, as follows from
(\ref{ahighentheta}). The crossing transformation
$s\leftrightarrow u$ of the amplitude (\ref{sumamplexp}) is
equivalent to the exchange $a(z)\leftrightarrow a(-z)\,$.
Therefore, the $t$-channel tree-level amplitude for the scattering
process $\phi\,\bar\phi\,\to\,\phi\,\bar\phi$ also goes to zero in
the ultraviolet if the analytic function $a(z)$ has another zero
at $z=+\infty\,$. Unfortunately, this is not possible if the power
series defining $a(z)\neq 0$ around zero is convergent on the
whole positive axis because all coefficients $a_r\geqslant 0$ of
the power series of $a(z)$ are non-negative. An interesting
possibility is therefore when the function $a(z)$ has a finite
radius of convergence around the origin. Outside the disk of
convergence, the function may be analytically continued and it is
this analytic continuation which determines the high energy
behaviour. A simple example is given by $a_r=r!$ in which case the
analytic continuation is given by $a(z)=(1-z)^{-1}$ which vanishes
for any large argument $z=\pm\infty\,$.\footnote{This example
exhibits a general feature: If the real function $a(z)$ defined by
the power series (\ref{pow ser}) with non-negative coefficients
$a_r\geqslant 0$ is meromorphic and has a finite radius of
convergence $R>0$ then $z=R$ is a pole of $a(z)$ on the positive
axis. The idea of the proof is as follows: The modulus of the
function $a$ at any given point $z_0$ inside the disk of
convergence satisfies the inequality $|a(z_0)|\leqslant \sum_r\,
\frac{|a_r|}{r!}\ |z_0|^r=\sum_r\, \frac{a_r}{r!}\
|z_0|^r=a(\,|z_0|\,)\,,$ because the coefficients $a_r$ are
non-negative. The function $a(z)$ is meromorphic and its power
series around the origin has a finite radius of convergence
$R>0\,$, thus it must have a pole $z_0$ on the circle of radius
$R\,$, \textit{i.e.} $|z_0|=R$ and $|a(z_0)|=\infty\,$. Therefore
$a(\,|z_0|\,)=a(R)=\infty\,$. In other words, $z=R$ is a
singularity of $a(z)$ on the positive axis, which can only be a
pole because the function is meromorphic.}

Consequently, the total scattering amplitude may be extremely soft
in the ultraviolet regime, though any individual exchange
amplitude grows quickly (for spin $r\geqslant 2$). Such asymptotic
behaviours may be qualitatively understood as follows: The
$t$-channel scattering amplitude of the process
$\phi\,\phi\,\to\,\phi\,\phi$ corresponding to an exchange of a
tensor field of rank $r$ behaves polynomially in the Regge limit
like $(-s)^{r}/t\,$, which is more and more divergent for larger
rank $r\,$. However, very precisely along the lines of \cite{GSW},
one may observe that the asymptotic behaviour of, say, the power
series $\sum_{r\geqslant 0} (-s)^r/r!=e^{-s}$ is much smaller when
$s\rightarrow+\infty$ than any individual term. Such a property
arises naturally for current-current interactions between two
scalar particles $\phi$ (or two scalar antiparticles $\bar{\phi}$)
because the series is alternating with the rank $r$ (remember that
the coefficients are non-negative $a_r\geqslant 0$).
Heuristically, some compensations are possible between the
exchanges of even-spin (attractive) and odd-spin (repulsive) gauge
tensors. Naively, this mechanism seems impossible between a scalar
particle $\phi$ and its antiparticle $\bar{\phi}$ or if the scalar
field is real ($\phi=\bar{\phi}$) because, intuitively, the former
interactions are always attractive. More precisely, the
$t$-channel scattering amplitude of the
$\phi\,\bar\phi\,\to\,\phi\,\bar{\phi}$ corresponding to an
exchange of a tensor field of rank $r$ behaves polynomially in the
Regge limit like $s^{r}/t\,$. Actually, the asymptotic behaviour
is a subtle issue because the non-compensation argument works only
inside the disk of convergence of the power series defining the
amplitude. Indeed, for instance the power series we already
mentioned $\sum_{r\geqslant 0} s^r=(1-s)^{-1}$ is not alternating
but its analytic continuation $(1-s)^{-1}$ goes to zero when
$s\rightarrow \infty\,$.

\subsection{Softness and finiteness}

The softness of tree-level scattering amplitudes in
the high-energy regime is a strong indication in favour of ultraviolet finiteness. 
For instance, various loop diagrams can be built out of the (off-shell) diagram 
of the previous section. As an illustration, one may consider the following 
one-loop contribution to the scalar propagator (encoding its self-energy)
\begin{fmffile}{self}
\begin{eqnarray}
\parbox{40mm}{
\begin{fmfgraph*}(40,25)
  \fmfleft{i}
  \fmfright{o}
  \fmf{fermion}{i,v1} \fmf{fermion}{v2,o}
  \fmf{photon,left,tension=.3}{v1,v2}
  \fmf{fermion}{v1,v2}
  \fmflabel{$k$}{i}
  \fmflabel{$k$}{o}
\end{fmfgraph*}}
\quad\quad\propto\quad \int d^4p\,\frac{A\Big(\phi(k)\phi(p)\to\phi(-p)\phi(k)\Big)\,}{p^2+m^2}\,,\nonumber
\end{eqnarray}
\end{fmffile}
where the internal curly lines should be understood
as the sum over all possible gauge fields and the amplitude $A$ is extended
off-shell.
This Feynman diagram can be seen to have at most a logarithmic divergence in the
UV if $a(z)$ goes to a constant when $z\rightarrow \pm\infty\,$. This is already
much better than any individual contribution coming from a finite number of gauge fields of spin $r\neq 0$ in the internal curly line.

Another example is the following box diagram which contributes to the two-scalar scattering process at one-loop.
\vspace{4mm}
\begin{fmffile}{simplebox}
\begin{eqnarray}
&&  \parbox{40mm}{
        \begin{fmfgraph*}(40,25)
            \fmfleft{i1,i2}
            \fmfright{o1,o2}
            \fmf{fermion}{i1,v1,v2,o1}
            \fmf{fermion,label}{i2,v3,v4,o2}
            \fmf{photon}{v1,v3}
            \fmf{photon}{v2,v4}
            \fmflabel{$k_1$}{i1}
            \fmflabel{$k_2$}{i2}
            \fmflabel{$\ell_1$}{o1}
            \fmflabel{$\ell_2$}{o2}
        \end{fmfgraph*}
}\quad\quad\propto\nonumber\\
\nonumber\\
\nonumber\\
&&\int d^4p\,\frac{A\Big(\phi(k_1)\phi(k_2)\to\phi(k_1+p)\phi(k_2-p)\Big)
\,A\Big(\phi(k_1+p)\phi(k_2-p)\to\phi(\ell_1)\phi(\ell_2)\Big)}{\Big((k_1+p)^2+m^2\Big)\,\Big((k_2-p)^2+m^2\Big)}\,,\nonumber
\end{eqnarray}
\end{fmffile}
This Feynman diagram can be seen to be UV finite if $a(z)$ goes to
some constant when $z\rightarrow \pm\infty$. Of course, this does
not imply that the corresponding total one-loop amplitudes are
finite because other diagrams should be taken into account, some
of which might include higher-order vertices which are not
considered in the present paper. Nevertheless, it is already very
suggestive to observe that some Feynman diagrams may be UV finite
if all contributions of the whole infinite tower of gauge fields
are summed.

\section{Non-relativistic interaction potential}\label{applications}

Since the higher-spin particles are massless, one may wonder about
the macroscopic interactions that they give rise to in $n=4$ dimensions.
In the low-energy (or non-relativistic) regime the Mandelstam variables
of the scattering process $\phi\,\phi\,\to\,\phi\,\phi$ examined in Section \ref{amplitudes}
behave as $s\sim 4m^2$ and $|u|\ll s\,$, thus the $t$-channel exchange amplitudes are equal to
\be
    \overset{(r)}{A}\big(\,\phi(\,\vec k\,)\,\phi(-\vec k\,)\,
    \to\,\phi(\,\vec\ell\ )\,\phi(-\vec\ell\ )\,\big)
    \sim\,-\,\frac{a_r}{r!}\ \Big(-\frac{(m\,\lambda)^2}2\Big)^{r-1}
    \frac{m^2}{(\,\vec k-\vec\ell\ )^2}\,.
    \label{amplow}
\ee
For the interaction arising from the exchange of a spin-$r$ mediator,
the non-relativistic potential between two elementary scalar particles
separated by $\vec x$ can be deduced from the above amplitude 
(\ref{amplow}) via the Born approximation (\ref{Born}) and reads
\be
\overset{(r)}{V}(\vec x)\,=\,\frac{a_r}{4\,r!}\,
\Big(-\frac{(m\lambda)^2}{2}\Big)^{r-1}\frac1{4\pi\, |\vec x|}\,,
\label{pot}
\ee
The sign indicates that even (odd) spin massless particles mediate
 attractive (repulsive) interactions between identical scalar particles (\textit{i.e.} charges of the same sign).
The effective non-relativistic potential including all possible exchanges is the sum
\begin{equation}
V(\vec x):=\sum\limits_{r\geqslant 0}\,\overset{(r)}{V}(\vec x)
\,=\,-\frac{1}{2\,(m\,\lambda)^2}\ a\Big(-\frac{(m\lambda)^2}{2}\Big)\,\frac1{4\pi\,|\vec x|}\,.
\label{effpot}
\end{equation}
This expression somehow justifies on physical ground the mathematical
 assumption that the function $a(z)$ should at least be analytic around zero. Indeed, in such case
$$
V(\vec x)\,=\,V_{\text{lower}}(\vec x)\,+\,{\cal O}\left(\,(m\lambda)^4\,\right)
$$
where $m\lambda\ll 1$ and $V_{\text{lower}}$ denotes the part of
 the effective potential corresponding to exchange of lower ($r\leqslant 2$)
  spin particles. In other words, the validity of the Taylor expansion of $a(z)$
  around zero agrees with the fact that higher-spin contributions are not
  observable at energy scales much smaller than Planck's mass.
In order to bring another perspective on this point, suppose now that we have two
 macroscopic bodies respectively made of $N\gg 1$ and $N^\prime\gg 1$ charged
 scalars (each of mass $m$). The macroscopic bodies have thus respective masses
 $M=N\,m$ and $M'=N'\,m\,$. The resulting macroscopic potential energy of the system
 for the interaction mediated by a massless spin-$r$ field is then obtained from (\ref{pot}) and reads
\be \overset{(r)}{W}(\vec x):=N\,N^\prime\,\overset{(r)}{V}(\vec
x)\,=
-\frac{\lambda^{2}}{8}\,\frac{a_r}{r!}\,\Big(-\frac{(m\lambda)^2}{2}\Big)^{r-2}\frac{M\,M'}{4\pi\,
|\vec x|}\,, \ee All macroscopic interaction potential can be
expressed in terms of the spin-two exchange (gravitational
interaction) as \be \overset{(r)}{W}(\vec x)\,=\,
\frac{2\,a_r}{r!\,a_2}\,\Big(-\frac{(m\lambda)^2}{2}\Big)^{r-2} \
\overset{(2)}{W}(\vec x)\,, \ee which clearly shows that when
${m\lambda}\ll 1$ (\textit{e.g.} if we take for $m$ the proton
mass and for $\lambda$ the inverse of the Planck mass) the
higher-spin interaction are negligible compared to the
gravitational ones. In order for the scalar exchange contribution
to be macroscopically invisible, one may assume that $a_0\ll
(m\lambda)^4\ll 1\,$. Moreover, if the macroscopic bodies are
approximately ``neutral'' (same number of particles and
antiparticles) then the odd-spin interactions are completely
negligible. The toy model considered here allows to understand why
higher-spin interactions would not be macroscopically observable
if they exist.

\section{Conclusion and discussion of results}\label{conclusion}

As advocated here, the Noether procedure applied to an infinite tower of (higher-rank)
conserved currents associated with (higher-derivative) symmetries of the Klein-Gordon
equation is deeply connected with Weyl quantisation and leads to a gauge symmetry
group which is (at lowest order) isomorphic to the group of unitary operators on ${\mathbb R}^n\,$.
In this picture, the scalar field transforms in the fundamental while the tower of
symmetric tensor gauge fields transforms in the adjoint representation of this group.
Apart from technical complications, the straight analogue of this cubic coupling between
a tower of (higher-spin) gauge fields and a free scalar field on any Riemannian manifold
$\mathcal M$ should lead to the group of unitary operators on $\mathcal M\,$.
The only difference would be that the Noether procedure could hold for homogeneous
manifolds only, in order for conserved currents to exist. Since only the simplest examples
of matter (a scalar field) and background (Minkowski spacetime) have been considered here,
the natural questions of how to extend the present analysis for spinor fields and/or
for constant-curvature spacetimes arise; they are currently under investigation.

The use of symbol calculus also
enables to write the cubic vertex in a very compact form which allows an explicit computation
of the general four-scalar tree-level amplitude. The coefficients of the exchanges of symmetric tensor gauge fields
 may be chosen in such a way that this amplitude is extremely soft in the high-energy regime.
For instance, the simplest choice of coefficients leads to an exponential fall-off of the
$\phi\,\phi\,\to\,\phi\,\phi$ high-energy tree amplitude which is very reminiscent of the behaviour
of the ultraviolet fixed-angle Veneziano/Virasoro four-tachyon amplitudes in open/closed string theory.
This suggestive property pleads in favour of the standard lore on higher-spin symmetries as
the deep origin of ultraviolet softness (and thereby maybe of perturbative finiteness) in string theory.
Further evidence in this direction would be provided by fixing the various coefficients
from some consistency requirement on the non-Abelian transformations in the gauge field sector.

At first sight, these non-trivial scattering amplitudes
and long-range interactions seem in contradiction with the various
 $S$-matrix no-go theorems on the interactions between matter
 and massless higher-spin particles \cite{Weinberg:1964ew,CMHLS}.
The main point is that the elastic scattering of  matter particles
is constrained to be trivial by higher-order conservation laws on
products of momenta, as in the case of free or even integrable
field theories. For instance, the conservation laws $\sum_i
k_i^{\mu_1}\ldots k_i^{\mu_r}=\sum_i \ell_i^{\mu_1}\ldots
\ell_i^{\mu_r}$ of order $r>1$ imply that the outgoing momenta can
only be a permutation of the incoming ones. On the one-hand, the
low-energy Weinberg theorem \cite{Weinberg:1964ew} states that
Lorentz invariance and the absence of unphysical degrees of
freedom from the amplitude of the
 emission of an external soft massless particle of spin $r$ imposes
  a conservation law of order $r\,$.
On the other hand, the conservation of higher-spin charges is associated with higher-order conservation laws, as in the Coleman-Mandula theorem \cite{CMHLS}. As a corollary, asymptotic higher-spin massless particles or conserved charges imply the triviality of the $S$-matrix. Like all theorems, the weakness of a no-go theorem relies in its assumptions.
In the present case, the fact that the scattering amplitudes of two scalars with some higher-spin field exchanged are non-trivial
could have several explanations, among which:
\begin{itemize}
 \item Asymptotic states of massless
 higher-spin particles may not exist in the complete theory,
 similarly to coloured states in QCD.
 \item It is necessary to fix the gauge in order to define the
 propagators for massless higher-spin fields, thus it is not obvious that their gauge symmetries automatically imply the existence of non-vanishing higher-spin conserved charges.
 \item The cubic vertex has been shown to be consistent at lowest order only,
 while the interactions might become inconsistent at higher-orders.
 \item There is no genuine $S$-matrix in (Anti) de Sitter space-time,
 so even if the cubic vertex is inconsistent in Minkowski space-time,
 its deformation in curved space-time might be consistent to all orders.
In a sense, the AdS/CFT correspondence is the definition of the
``$S$-matrix'' in Anti de Sitter space-time  \cite{Witten98}.
Therefore, an infinite number of asymptotic higher-spin conserved
charges means that the holographic dual theory is integrable, but
it does not imply that the ``scattering'' theory in the bulk
(defined by the Witten diagrams) is trivial at all. This
observation is indeed the very basis of the holographic
correspondence in the higher-spin context \cite{Su}.
  \item Along these lines, another possibility is that, when $m=0\,$, 
  the action $\bra\phi \hat G \ket\phi$ could be interpreted as the action for 
  a conformal scalar field $\phi$ living on the ``boundary'' of AdS
  and interacting with higher-spin gauge fields in the ``bulk'' 
  (see \cite{Segal:2000ke} for similar line of reasoning).
\end{itemize}

To end up, the  issue of the trace constraints of Fr\o nsdal \cite{Fronsdal} on the gauge fields
and parameters in higher-spin metric-like theory has not been discussed in the previous sections
and deserves some comments. These constraints might have been included by consistently imposing
weaker conservation laws on double-traceless currents. This would not modify the current-current interactions
because the residue of the propagator is automatically double-traceless, as pointed out in \cite{FMS}.
Nevertheless, it was convenient to remove trace constraints when reflecting on the non-Abelian symmetry group.
Anyway, the trace constraints may be removed in the action principle for free higher-spin metric-like fields
in several ways (see \cite{FS} for some reviews, and \cite{FMS,BGK} for some later developments).
As far as the non-Abelian frame-like formulation is concerned, the analogues of
Vasiliev's unfolded equations in the unconstrained case \cite{G,V} are dynamically empty and can somehow
be thought \cite{G,V,Grigoriev} of as Fedosov's quantisation \cite{Fedosov} of the cotangent bundle along the lines of \cite{Bordemann:1997ep}. But a slight refinement of Vasiliev's unfolded equations \cite{unfold} has been proposed in \cite{SSS} and should also be dynamically interesting. The frame-like formalism with weaker trace constraints \cite{Sorokin:2008tf} might also prove to be useful in this respect.
Last but not least, the group of gauge symmetries of the metric-like theory arising from
unconstrained frame-like theories (by fixing the gauge and solving the torsion constraints)
can be shown to be also isomorphic to the group of unitary operators on ${\mathbb R}^n\,$,
at lowest order in the gauge fields and around flat spacetime \cite{XB}.

\acknowledgments{ We thank D. Francia  and A. Sagnotti for many helpful discussions.
X.B. is grateful to the organisers of the ``6th International
Spring School and Workshop on Quantum Field Theory and Hamiltonian
Systems'' (Calimanesti \& Caciulata, Romania\,; May 2008) for
their invitation to an enjoyable meeting and the opportunity to
present lectures (published in \cite{BJM}) on related results.
X.B. also acknowledges N. Boulanger and P. Sundell for general discussions, S. Nicolis and O. Thibault for their help in drawing Feynman diagrams with {\tt feynMF} package.
}

\appendix

\section{Weyl quantisation}\label{Weylquantization}

The Weyl formalism \cite{WW} offers a classical-like formulation of quantum
mechanics
using phase space functions as observables and the Wigner function as an
analogue of the Liouville density function.

In order to fix the ideas, one may consider the simplest case: the
quantum description of a single particle. Classical mechanics is
based on the commutative algebra of classical observables
(\textit{i.e.} real functions $f(x^\mu,p_\nu)$ on the phase space
$T^*{\mathbb R}^n\cong{\mathbb R}^{n}\times{\mathbb R}^{n*}$)
endowed with the canonical Poisson bracket
$$
\{f,g\}_{\textbf{P.B.}}\,=\,\frac{\partial f}{\partial x^{\mu}}\,\frac{\partial
g}{\partial p_{\mu}}\, -\,\frac{\partial f}{\partial
p_{\mu}}\,\frac{\partial g}{\partial x^{\mu}}\,.
$$
The \textit{Weyl map} ${\cal W}:f(x^\mu,p_\nu)\mapsto \hat{F}$
associates to any function $f$ a Weyl(\textit{i.e.}
symmetric)-ordered operator $\hat{F}$ defined by
\begin{equation}
\hat{F}\,=\,\frac{1}{(2\pi\hbar)^{n}}\int d^nk\,d^nv\,\,{\cal
F}(k,v)\,e^{\frac{i}{\hbar}\,(\,k_\mu\,\hat{X}^\mu\,-\,v^\mu\,\hat{P}_\mu)}\,,
\label{Weylmap}
\end{equation}
where ${\cal F}$ is the Fourier transform\footnote{The Weyl map is
well defined for a much larger class than square integrable
functions, including for instance the polynomial functions
(remark: their Fourier transform are distributions).} of $f$ over
\textit{whole} phase space (in other words, over position
\textit{and} momentum spaces)
$$
{\cal F}(k,v)\,:=\,\frac{1}{(2\pi\hbar)^{n}}\int
d^nx\,d^np\,\,f(x,p)\,e^{-\frac{i}{\hbar}\,(\,k_\mu\,x^\mu\,-\,v^\mu\,p_\mu)}\,.
$$
The function $f(x,p)$ is called the \textit{Weyl symbol} of the
operator $\hat{F}\,$, which need not be in symmetric-ordered form.
A nice property of the Weyl map (\ref{Weylmap}) is that it relates
the complex conjugation $^*$ of symbols to the Hermitian
conjugation $^\dagger$ of operators, ${\cal
W}:f^*(x^\mu,p_\nu)\mapsto \hat{F}^\dagger$. Consequently, the
image of a real function (a classical observable) is a Hermitian
operator (a quantum observable). The inverse ${\cal
W}^{-1}:\hat{F}\mapsto f(x^\mu,p_\nu)$ of the Weyl map is called
the \textit{Wigner map}.

The commutation relations between the position and momentum
operators are
$[\hat{X}^\mu,\hat{P}_\nu]_{_-}=i\,\hbar\,\delta^\mu_\nu\,$, where
$[\hat{A},\hat{B}]_\pm:=\hat{A}\hat{B}\pm\hat{B}\hat{A}$ denotes
the (anti)commutator of the operators $\hat{A}$ and $\hat{B}\,$.
The Baker-Campbell-Hausdorff formula implies that if the
commutator $[\hat{A},\hat{B}]_{_-}$ itself commutes with both
$\hat{A}$ and $\hat{B}\,$, then
$$
e^{\hat{A}}\,e^{\hat{B}}=e^{\hat{A}\,+\,\hat{B}\,+\,\frac12\,[\hat{A},\hat{B}]_{_-}}\,.
$$
Moreover, for any operators $\hat{A}$ and $\hat{B}$ one can show
that
$$
e^{\hat{A}}\,e^{\hat{B}}\,e^{{\pm}\hat{A}}=e^{[\hat{A},\,\,\,]_\pm}\,e^{\hat{B}}\,\,,
$$
where $[\hat{A},\,\,\,]_{_\pm}$ denotes the (anti)adjoint action
of $\hat{A}\,$. Two very useful equalities follow:
\begin{eqnarray}
\,e^{\frac{i}{\hbar}\,(\,k_\mu\,\hat{X}^\mu\,-\,v^\mu\,\hat{P}_\mu)}
&=&e^{\frac{i}{2\hbar}\,k_\mu\,\hat{X}^\mu}\,e^{-\frac{i}{\hbar}\,
v^\mu\,\hat{P}_\mu}\,e^{\frac{i}{2\hbar}\,k_\mu\,\hat{X}^\mu}\label{usefind}\\
&=&e^{\frac{i}{2\hbar}\,k_\mu\,[\,\hat{X}^\mu,\,\,\,]_+}\,e^{-\frac{i}{\hbar}\,v^\mu\,\hat{P}_\mu}
\label{usefind2}
\end{eqnarray}
Combining (\ref{Weylmap}) with (\ref{usefind2}) implies that one
way to explicitly perform the Weyl map is via some
``anticommutator ordering'' for half of the variables with respect
to their conjugates.

The matrix elements in the position basis of the exponential
operator in (\ref{Weylmap}) are found to be equal to
\begin{eqnarray}
&&\langle\,x\mid e^{\frac{i}{\hbar}\,(\,k_\mu\,\hat{X}^\mu\,-\,v^\mu\,\hat{P}_\mu)}\mid x^\prime\,\rangle \nonumber\\
&&\quad=\,e^{\frac{i}{2\hbar}\,k_\mu (\,x^\mu+\,x^{\prime\,\mu})}\,\langle\,x\mid e^{-\frac{i}{\hbar}\,v^\mu\,\hat{P}_\mu}\mid x^\prime\,\rangle\nonumber\\
&&\quad=\int \frac{d^np}{(2\pi\hbar)^{n}}\,e^{\frac{i}{2\hbar}\,k_\mu
(\,x^\mu+\,x^{\prime\,\mu})\,+\,\frac{i}{\hbar}\,(\,x^\mu-x^{\prime\,\mu}-\,v^\mu)\,p_\mu}
\label{previousrel}
\end{eqnarray}
by making use of the identity (\ref{usefind}) and by inserting the
completeness relation $\int d^np/(2\pi\hbar)^n\,\mid p\,\rangle\,\langle\,
p\mid\,=\widehat{\,1}\,$.

The \textit{integral kernel} of an operator $\hat{F}$ is the
matrix element $\langle\,x\mid\hat{F}\mid x^\prime\rangle$
appearing in the position representation of the state $\hat{F}\mid
\psi\,\rangle$ as follows
$$
\langle\, x\mid\hat{F}\mid \psi\,\rangle\,=\,\int d^nx^\prime
\,\,\psi(x^\prime)\,\, \langle\, x\mid\hat{F}\mid
x^\prime\,\rangle\,,
$$
where the wave function in position space is
$\psi(x^\prime):=\langle\, x^\prime\mid\psi\,\rangle$ and the
completeness relation $\int dx^\prime\mid
x^\prime\,\rangle\,\langle\, x^\prime\mid\,=\widehat{\,1}$ has
been inserted. The definition (\ref{Weylmap}) and the previous
relation (\ref{previousrel}) enable to write the integral kernel
of an operator in terms of its Weyl symbol,
\begin{equation}
\langle\,x\mid\hat{F}\mid
x^\prime\,\rangle\,=\,\int\frac{d^np}{(2\pi\hbar)^{n}}\  f\,\big(\,\frac{x+x^\prime}{2}\,,\,p\,\big)\,\,e^{\frac{i}{\hbar}\,(\,x^\mu-x^{\prime\,\mu})\,p_\mu}\,.
\label{matrixelement}
\end{equation}
This provides an explicit form of the Wigner map
\begin{equation}
f(x^\mu,p_\nu)\,=\,\int d^nq\,\,\langle\,x-q/2\mid\hat{F}\mid
x+q/2\,\rangle\,\,e^{\frac{i}{\hbar}\, q^\mu\,p_\mu}\,,
\label{Wignermap}
\end{equation}
as follows from the expression (\ref{matrixelement}). This shows
that indeed the Weyl and Wigner maps are bijections between the
vector spaces of classical and quantum observables. The Fourier
transform
$$
\breve{f}(x^\mu,v^\nu)\,:=\int \frac{d^np}{(2\pi\hbar)^{n}}\
f(x^\mu,p_\nu)\,\,e^{\frac{i}{\hbar}\, v^\mu\,p_\mu}\,,
$$
over momentum space of the Weyl symbol $f(x,p)$ is a function on
the configuration space $T{\mathbb R}^{n}\cong {\mathbb
R}^{2n}\,$. The equation (\ref{matrixelement}) states that the
Fourier transform over momentum space of the Weyl symbol is
related to the integral kernel of its operator via
\begin{equation}
\langle\,x\mid\hat{F}\mid
x^\prime\,\rangle\,=\,\breve{f}\,\big(\,\frac{x+x^\prime}{2}\,,\,x^\mu-x^{\prime\,\mu}\,\big)
\label{intkernFtr}
\end{equation}
or, equivalently,
\begin{equation}
\breve{f}\big(x^\mu,v^\nu\big)\,=\,\langle\,x+v/2\mid\hat{F}\mid
x-v/2\,\rangle\,. \label{Fouriertransform}
\end{equation}
By integrating over $x=x^\prime\,$, the relation
(\ref{matrixelement}) also implies that the trace of an operator
$\hat{F}$ is proportional to the integral over phase space of its
Weyl symbol $f\,$,
\begin{equation}
\mbox{Tr}[\hat{F}]\,=\,\frac{1}{(2\pi\hbar)^{n}}\int
d^nx\,d^np\,\,f(x,p)\,. \label{trace}
\end{equation}
As a side remark, notice that the Fourier transform
$$
\tilde{f}(k_\mu,p_\nu)\,:=\,
\,\int
d^nx\,\,f(x^\mu,p_\nu)\,\,e^{-\,\frac{i}{\hbar}\, k_\mu\,x^\mu}\,,
$$
over position space of the Weyl symbol $f(x,p)$ is related to the
matrix element in the momentum basis of the operator $\hat{F}$ via
\begin{equation}
\langle\,k\mid\hat{F}\mid
k^\prime\,\rangle\,=\,\tilde{f}\,\big(\,k^\mu-k^{\prime\,\mu}\,,\,\frac{k+k^\prime}{2}\,\big)
\label{intkernFtr2}
\end{equation}
in direct analogy with (\ref{intkernFtr}).

The \textit{Moyal product} $\star$ is the pull-back of the
composition product in the algebra of quantum observables with
respect to the Weyl map $\cal W\,$, such that the latter becomes
an isomorphism of associative algebras, namely
\begin{equation}
{\cal W}\big[f(x,p)\,\star\, g(x,p)\big]\,=\,\hat{F}\,\hat{G}\,.
\label{morphism}
\end{equation}
The Wigner map (\ref{Wignermap}) allows to check that the
following explicit expression of the Moyal product satisfies the
definition (\ref{morphism}),
\begin{eqnarray}
f(x,p)\,\star\, g(x,p)&=&f(x,p)\,\,\exp\left[\,\frac{i\,\hbar}{2}\,\left(\frac{\overleftarrow{\partial}}{\partial x^{\mu}}\,\frac{\overrightarrow{\partial}}{\partial p_{\mu}}-\frac{\overleftarrow{\partial}}{\partial p_{\mu}}\,\frac{\overrightarrow{\partial}}{\partial x^{\mu}}\right)\right]\,g(x,p)\nonumber\\
&=&f(x,p)\,
g(x,p)+\frac{i\,\hbar}{2}\left\{f(x,p)\,,\,g(x,p)\right\}_{\textbf{P.B.}}+{\cal
O}(\hbar^2) \label{Moyalproduct}
\end{eqnarray}
where the arrows indicate on which factor the derivatives should
act. The trace formula (\ref{trace}) for a product of operators
leads to
\begin{eqnarray}
\mbox{Tr}[\,\hat{F}\hat{G}\,]&=&\frac{1}{(2\pi\hbar)^{n}}\int dx\,dp\,\,f(x,p)\,\star\, g(x,p)\nonumber\\
&=&\frac{1}{(2\pi\hbar)^{n}}\int dx\,dp\,\,f(x,p)\, g(x,p)
\label{traceproduct}
\end{eqnarray}
because all terms in the Moyal product (\ref{Moyalproduct}) beyond
the pointwise product are divergences over phase space and any
boundary term will always be assumed to be zero in the present
notes.

The \textit{Wigner function} $\rho(x,p)$ is the Weyl symbol of the
density operator $\hat{\rho}$ under the Wigner map
(\ref{Wignermap}). Let $\mid\psi\,\rangle$ be an (unnormalised)
quantum state. The corresponding pure state density operator is
equal to $\hat{\rho}\,:=\,\mid\psi\,\rangle\langle\,\psi\mid\,$.
Then the Fourier transform over momentum space of the pure state
Wigner function $\rho(x,p)$ can be written in terms of the wave
function $\psi(x)$ as follows,
\begin{equation}
\breve{\rho}(x,q)=\psi(x+q/2)\,\psi^*(x-q/2)\,, \label{rhotilde}
\end{equation}
due to (\ref{Fouriertransform}). The mean value of an observable
$\hat{F}$ over the state $\mid\psi\,\rangle$ is proportional to
the integral over phase space of the product between the Wigner
function $\rho$ and the Weyl symbol $f\,$,
\begin{equation}
\langle\,F\,\rangle_{\,\psi}\,=\,\frac{\langle\,\psi\mid\hat{F}\mid\psi\rangle}{\langle\,\psi\mid\psi\rangle}\,
=\,\frac{\mbox{Tr}[\,\hat{\rho}\,\hat{F}\,]}{\mbox{Tr}\,[\,\hat{\rho}\,]}\,
=\,\frac{\int dx\,dp\,\,\rho(x,p)\, f(x,p)}{\int
dx\,dp\,\,\rho(x,p)}\,, \label{quasiproba}
\end{equation}
which explains why the Wigner function is sometimes called the
Wigner ``quasi-probability distribution.'' It should be emphasised
that the Wigner function is real but may take negative values,
thereby exhibiting quantum behaviour.

Let $\hat{H}$ be a Hamiltonian operator of Weyl symbol $h(x,p)\,$.
In the Heisenberg picture, the time evolution of quantum
observables (which do not depend explicitly on time) is governed
by the differential equation
\begin{equation}
\frac{d\hat{F}}{dt}\,=\,\frac{1}{i\,\hbar}\,[\hat{F},\hat{H}]_{_-}\quad\Longleftrightarrow\quad
\frac{df}{dt}\,=\,\frac{1}{i\,\hbar}\,[\,f\overset{\star}{,}h\,]_{_-}
\label{timevolution}
\end{equation}
where $[\,\,\,\overset{\star}{,}\,\,\,]_{_-}$ denotes the
\textit{Moyal commutator} defined by
\begin{eqnarray}
&&\left[\,f(x,p)\,\overset{\star}{,}\, g(x,p)\,\right]_{_-}:=f(x,p)\,\star\, g(x,p)\,-\,g(x,p)\,\star\, f(x,p)\nonumber\\
&&\quad=\,2\,i\,\,f(x,p)\,\,\sin\left[\,\frac{\hbar}{2}\,\left(\frac{\overleftarrow{\partial}}{\partial x^{\mu}}\,\frac{\overrightarrow{\partial}}{\partial p_{\mu}}-\frac{\overleftarrow{\partial}}{\partial p_{\mu}}\,\frac{\overrightarrow{\partial}}{\partial x^{\mu}}\right)\right]\,g(x,p)\nonumber\\
&&\quad=\,i\,\hbar\,\left\{\,f(x,p)\,,\,g(x,p)\,\right\}_{\textbf{P.B.}}\,+\,{\cal
O}(\hbar^2)\,, \label{Moyalcommutator}
\end{eqnarray}
as can be seen from (\ref{Moyalproduct}). essentially to the
Poisson bracket. The \textit{Moyal bracket} is the renormalisation
of the Moyal commutator given by
$$\frac{1}{i\,\hbar}\,[\,\,\,\,\overset{\star}{,}\,\,\,\,]_{_-}
=\left\{\,\,\,,\,\,\right\}_{\textbf{P.B.}}+{\cal O}(\hbar).\,$$
The Moyal bracket is a deformation of the Poisson
bracket, and one can see that the equation (\ref{timevolution}) in
terms of the Weyl symbol is a perturbation of the Hamiltonian
flow. If either $f(x,p)$ or $g(x,p)$ is a polynomial of degree
two, then their Moyal bracket reduces to their Poisson bracket. So
when the Hamiltonian is quadratic (free) the quantum evolution of
a Weyl symbol is identical to its classical evolution.

\section{Elastic scattering}\label{mandelstam}

The three \textit{Mandelstam variables} $s\,$, $t$ and $u$
of any elastic scattering of four particles (see \textit{e.g.} the textbook \cite{Hagedorn}) with the same mass $m$ are related by
$s+t+u=4\,m^2\,$. In $n=4$ dimensions,
there are indeed only two independent Lorentz invariants which can be constructed from the four $4$-momenta.

Let the Mandelstam variables of the scattering $\phi(k_1)\,\phi(k_2)\,\to\,\phi(\ell_1)\,\phi(\ell_2)$ be
\begin{equation}
s=-(k_1+k_2)^2\,,\quad t=-(k_1-\ell_1)^2\,,\quad u=-(k_1-\ell_2)^2\,.
\label{Mandelstam}
\end{equation}
In the center-of-mass system, the four-momenta take the form
$$
k^\mu_1=\left(\frac{\sqrt{s}}2\,,\,{\vec k}\right)\,,\quad k^\mu_2=\left(\frac{\sqrt{s}}2\,,\,-{\vec k}\right)\,,\quad \ell^\mu_1=\left(\frac{\sqrt{s}}2\,,\,{\vec \ell}\right)\,,\quad \ell^\mu_2=\left(\frac{\sqrt{s}}2\,,\,-{\vec \ell}\right)\,,
$$
hence the variable $s\geqslant (2m)^2$ is the \textit{squared center of mass energy}, the variable
$t=-({\vec k}-{\vec \ell}\,)^2$ is the \textit{squared momentum transfer} and $u=-({\vec k}+{\vec \ell}\,)^2$ has no obvious physical interpretation.
The (center-of-mass) \textit{scattering angle} $\theta$ is defined as the angle between $\vec k$ and $\vec\ell\,$.
The products of momenta are related by
$$
k_1\cdot k_2=\ell_1\cdot \ell_2= m^2-\frac{s}{2}\,,\,\quad k_1\cdot \ell_1=k_2\cdot \ell_2=\frac{t}{2}-m^2 \,,
$$
$$
k_1\cdot \ell_2=k_2\cdot \ell_1=\frac{u}{2}-m^2 \,.
$$
The two relevant variables of the problem considered in the paper are
$$s+u=-(k_1+\ell_1)^2=-(k_2+\ell_2)^2\,,\quad s-u=-(k_1+\ell_1)\cdot(k_2+\ell_2)\,.$$
Both variables can be expressed in terms of the squared center-of-mass energy $s$ and scattering angle $\theta$ as
$$s\pm u=\left[\,1\mp \cos^2\left(\frac{\,\theta\,}{2}\right)\right]\,s\,\pm 4m^2\,.$$
Hence, for large $s\gg m^2\,$, they behave in the Regge limit\footnote{
    Traditionnally, one distinguishes two high-energy
    (\textit{i.e.} $s/m^2\rightarrow\infty$) limits:
    the \textit{Regge} (or \textit{fixed momentum transfer}) limit
    which corresponds to $s/m^2\rightarrow\infty$ with $t$ fixed
    (thus $\theta\rightarrow 0$ and $u/m^2\rightarrow -\infty$)
    and the \textit{fixed scattering angle} limit which corresponds to
    $s/m^2\rightarrow\infty$ with $s/\,t$ and $u/\,s$ fixed
    (thus $t/m^2\rightarrow -\infty$ and $u/m^2\rightarrow -\infty$).} as
$$s+u\,\sim\, -\,t\quad\mbox{is fixed and}\quad\frac{s-u}{s+u}\,\sim\,-\,\frac{2}{t}\,s\quad\mbox{is large}\,,$$
and in the fixed scattering angle limit as
$$\frac{s-u}{s+u}\sim\frac{1+\cos^2\left(\frac{\,\theta\,}{2}\right)}{\sin^2\left(\frac{\,\theta\,}{2}\right)}\quad\mbox{is fixed and}\quad s+u\sim \sin^2\left(\frac{\,\theta\,}{2}\right)\,s\quad\mbox{is large}\,.$$

\bigskip

In the scattering theory of quantum mechanics, the differential cross section
between two boson is given by
\be
    \frac{d\,\sigma}{d\,\Omega}=\left|\,f(\vec k\,,\vec \ell\ )+f(\vec k\,,-\vec \ell\ )\,\right|^2\,.
    \label{cs qm}
\ee
and in the \textit{Born approximation} the scattered waves $f$ are proportional
to the Fourier transform of the potential $V(\vec x\,)\,$:
\be
f(\vec k\,,\vec \ell\ )=-\frac{m}{4\pi}\int d^3 x \,V(\vec x\,)\,e^{i\,(\vec k\,-\vec \ell\ )\cdot \vec x}\,.
\label{Born}
\ee
Comparing eq.(\ref{cs qm}) with the differential cross section calculated from
the scattering amplitude $A(s,t,u)$\,:
\be
    \frac{d\,\sigma}{d\,\Omega}=\frac{1}{64\,\pi^2\,s}
    \,\left|\,A(s,t,u)+A(s,u,t)\,\right|^2\,,
\ee
we get the non-relativistic interaction potential as
\be
    V(\vec x\,)=-\frac1{4\,m^2}\int \frac{d^3p}{(2\pi)^3}\ A\big(4m^2,-{\vec p\, }^2,0\big)\ e^{-i\,\vec p\,\cdot \vec x}\,.
\ee

Remark that another way\footnote{See \textit{e.g.} \cite{Jagannathan:1986hr} for the case of massive mediating fields.} of deriving the low-energy interaction potentials is by considering a distribution
\begin{equation}
\ru J{\mu_1\ldots\mu_r}(x)\,=\,\sigma(x)\,\,w^{\mu_1}\ldots w^{\mu_r}
\label{explcurrents}
\end{equation}
of particles at rest of density and velocity respectively given by
the scalar $\sigma(\vec x)$ and the fixed vector $w^\mu\,$.
Plugging (\ref{explcurrents}) inside the integrals (\ref{fmsfin})
leads to a current-current interaction in $n=4$ given by
\begin{eqnarray}
\overset{(r)}{S}_{\text{curr}}\,[\,\sigma\,]&=&\int dx^0\,\int d^3x\,\,\sigma(\vec x)\,\,\frac{1}{\Delta}\,\,\sigma(\vec x)\,,\nonumber\\
&=&\int dx^0\,\int d^3x\,d^3y\,\,\sigma(\vec x)\,\,\sigma(\vec y)\,\,\overset{(r)}{V}(\rho)\,,
\end{eqnarray}
where $\overset{(r)}{V}(\rho)$ is the interaction potential computed in (\ref{pot}) where $m\lambda=1\,$.

\section{Chebyshev and Gegenbauer polynomials}\label{Gegenbauer}

Several useful definitions and formulas taken from \cite{orthopols} are collected here in order to be self-contained.

The \textit{Chebyshev polynomial of first kind} $T_r(z)$ is uniquely defined by the relation $T_r(\cos\beta) = \cos(r\beta)$ for any angle $\beta$, which implies
\be
T_r(z)=\frac12\left[\left(z+\sqrt{z^2-1}\right)^r+\left(z-\sqrt{z^2-1}\right)^r\,\right]
\label{Chebypower}
\ee
 but it is also given by the sum
\begin{equation}
T_r(z)\,=\,\frac{r}2\sum\limits^{[\frac{r}2]}_{m= 0}\,\frac{(-1)^m (r-m-1)!}{\,m!\,(r-2m)!}\, (2z)^{r-2m}\,,
\label{Tdefinition}
\end{equation}
when $r\geqslant 1\,$. Observe that $T_0(z)=1=T_r(1)$ and $T_r(-z)=(-1)^r T_r(z)\,$. When $|z|\gg 1\,$, the Chebyshev polynomial of first kind with index $r\geqslant 1$ behaves as $T_r(z)\sim 2^{r-1}\,r\,z^r$.

The \textit{Gegenbauer (or ultraspherical) polynomial} $C^\alpha_r(z)$ with $\alpha>-\frac12$ et $\alpha\neq 0$ is a polynomial of degree $r\in\mathbb N$ in the variable $z$ defined as
\begin{equation}
C^{\alpha}_r(z)\,:=\,\sum\limits^{[\frac{r}2]}_{m= 0}\,\frac{(-1)^m (\alpha)_{r-m}}{\,m!\,(r-2m)!}\, (2z)^{r-2m}\,.
\label{Gdefinition}
\end{equation}
They generalize the Legendre polynomials $P_r(z)$ to which $C^\frac12_r(z)$ is proportional. Moreover, the Chebyshev polynomial of first kind $T_r(z)$ may somehow be thought as a regularised limit of Gegenbauer polynomials $C^{\alpha}_r(z)$ for $\alpha\rightarrow 0\,$.
Notice that $C_0^\alpha(z)=1$ and $C^\alpha_r(-z)=(-1)^r C^\alpha_r(z)\,$. When $|z|\gg 1\,$, the Gegenbauer polynomial behaves as $C^{\alpha}_r(z)\sim (\alpha)_r(2z)^r/r!\,\,$.

\end{document}